\documentclass[journal,twoside,twocolumn]{IEEEtran}

\usepackage{amsmath,amssymb,amsfonts}
\usepackage{bm}
\usepackage{graphicx}
\usepackage{cite}
\usepackage{url}
\usepackage{booktabs}
\usepackage{array}
\usepackage{multirow}
\usepackage{xcolor}
\usepackage{algorithm}
\usepackage{hyperref}
\usepackage{orcidlink}
\usepackage{amsthm}
\hypersetup{colorlinks=false}
\usepackage{algorithmicx}
\usepackage{algpseudocode}
\usepackage{bbm}
\usepackage{enumitem}

\begin{document}

\title{Angularly Robust Constellation Synthesis and Calibration of Oversampling Digital-to-Phase Transmit Arrays}

\author{
  Anahita~Asadi \orcidlink{0009-0007-6133-9998},
  Justin~J.~Kim \orcidlink{0009-0007-3203-7165},
  James~F.~Buckwalter \orcidlink{0000-0002-9390-0897},
  and~Inna~Partin-Vaisband \orcidlink{0000-0002-6399-6672}

\thanks{A.~Asadi and I.~Partin-Vaisband are with the Department of Electrical and Computer Engineering, University of Illinois Chicago, Chicago, IL 60607 USA (email: aasadi5@uic.edu; vaisband@uic.edu).}
\thanks{J.~J.~Kim and J.~F.~Buckwalter are with the Department of Electrical and Computer Engineering, University of California Santa Barbara, Santa Barbara, CA 93106 USA (email: jjkim@ucsb.edu; buckwalter@ece.ucsb.edu).}
\thanks{This work was supported in part by the CogniSense: Center on Cognitive Multi-spectral Sensors, one of seven centers in Joint University Microelectronics Program (JUMP) 2.0, a Semiconductor Research Corporation (SRC) program sponsored by the Defense Advanced Research Projects Agency (DARPA).}
}

\maketitle

\begin{abstract}
Oversampling digital-to-phase transmitters replace per-channel digital-to-analog converters with 2-bit phase shifters and synthesize amplitude through spatial and temporal combining.
Codeword selection over the $(2^b)^{N \times T}$ configurations for $b$-bit phase states, $N$ channels, $T$ temporal subslots must handle array interference, hardware imbalance, and pointing drift.
This work formulates constellation synthesis over per-channel symbol-averaged excitations with phase-state and channel responses. Under a second-order Taylor approximation, the expected error under Gaussian pointing drift is a Hermitian positive-semidefinite quadratic of rank at most three.
At 512 channels without drift, randomization raises greedy sidelobe suppression from 7.85 to 12.34~dB at an unchanged EVM of 4.51\%, within 0.7~dB of simulated annealing with cached response sums at 13--16$\times$ lower runtime.
In ideal-hardware 16-QAM simulations, drift-aware design reduces EVM under drift by up to 1.54$\times$.
On a 4-channel 22-nm FD-SOI array, phase-state gains fitted away from a held-out angle and its neighbors let 16 pilots predict its other 240 responses with mean errors of 5.46\% and 5.32\% for 16- and 32-QAM, up to 13.1$\times$ below an ideal-response model.
Codebooks selected from these predictions measure within 1.7 percentage points of selection from all 256 measured responses, and drift-aware synthesis reduces measured EVM under drift by up to 1.18$\times$.
\end{abstract}


\section{Introduction}
\label{sec:intro}

\IEEEPARstart{P}{hased} arrays provide the directional gain needed for communication and sensing at D-band frequencies (110--170~GHz) \cite{farid2022tmtt}. 
In a conventional transmit (TX) array, each power amplifier (PA) must preserve the amplitude variations of a quadrature amplitude modulation (QAM) signal.
Constant-envelope signals let each PA run saturated and efficient, whereas QAM forces linear and inefficient PAs to avoid distortion. Constant-envelope arrays instead modulate amplitude through the relative phases of the channels combined at the receiver \cite{liang2011jssc,mohammed2012twc}.
However, analog phase shifters (PS) demand tight per-channel calibration across hundreds of channels \cite{li2022jssc}.
Fully digital arrays, on the other hand, require power-hungry high-speed digital to analog converters (DACs) on every channel \cite{wang2018twc}.
Therefore, achieving high effective isotropic radiated power (EIRP) directly increases power consumption and hardware costs. 

To reduce these costs, QAM is instead performed through \textit{precoding} (i.e., jointly selecting the signals transmitted by multiple channels so that their combination at the receiver produces the desired QAM symbol) with some quantization error.
For this, $\Sigma\Delta$ precoding uses spatial feedback across discrete-phase elements \cite{shao2019jstsp}, whereas quadrature phase-shift keying (QPSK) superposition \cite{oveisi2023tcas} and multilevel outphasing arrays \cite{lampu2023tcomm} directly combine constant-envelope, phase-controlled signals.
Oversampling digital to phase (ODP) TX arrays additionally utilize temporal aggregation to enrich constellations \cite{kim2025rfic}.
For ODP TX arrays, the broadside constellation count and a quantization-limited EVM expression for rotated 16-QAM are derived, off-broadside constellations are shown, pilot-assisted calibration is performed at individual angles, and oversampled 16- and 32-QAM waveforms are measured \cite{kim2026jmw}.
Transferring a calibrated model across angles, predicting unmeasured codewords at a new angle from a few pilots, and selecting codewords under angular uncertainty remains open.

\begin{figure}
    \centering
    \includegraphics[width=0.95\linewidth]{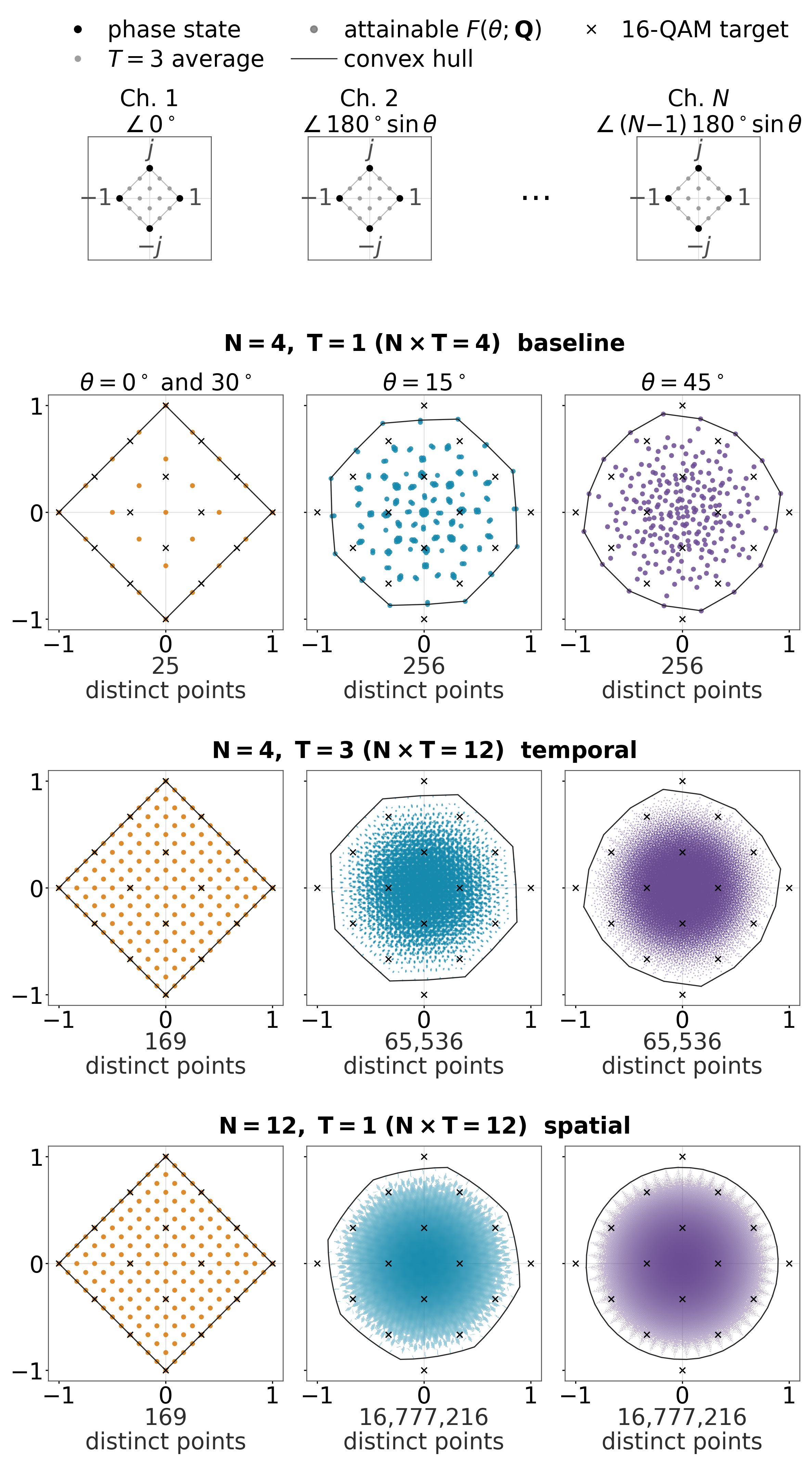}
    \caption{Effect of temporal and spatial sampling on reachable constellation states at different angles with half-wavelength spacing. 
    Top: each channel's four phase states and the averages attainable with $T=3$. 
    Bottom: attainable array factors (dots), their convex hulls (outlines), and ideal 16-QAM targets (crosses) at $0^\circ$, $15^\circ$, and $45^\circ$. The channel count $N$, subslot count $T$, and product $NT$ are indicated for each configuration.}
    \label{fig:odp}
\end{figure}

The ODP approach, however, shifts the burden from hardware to algorithm. 
The transmitted symbol is synthesized through the simultaneous over-the-air superposition of all array channels and their collective phase configuration.
For an $N$-channel array, each with $b$-bits and sampled in $T$ equal-duration subslots, the array can be configured in $(2^b)^{N \times T}$ ways. 
Fig.~\ref{fig:odp} shows how quickly the resulting constellation grows as $N$ and $T$ grow. 
Exhaustive search over configurations becomes intractable; nevertheless, searching the space is necessary for a high-quality design.

Each $N\times T$ matrix of phase states is called a \textit{codeword}, and the set of $(2^b)^{NT}$ codewords for all constellation symbols at a specified steering angle is a \textit{codeword space}.
Transmitting an $M$-point constellation then requires a \textit{codebook}, i.e., $M$ codewords whose array factors best approximate the $M$ target points.
Selecting such a codebook from an exponentially large codeword space, while remaining robust to hardware perturbations of ODP TX arrays, requires an efficient method.

The closest algorithmic approaches to solve this problem are as follows:
\begin{itemize}
    \item \textbf{Discrete-phase and finite-codebook precoding.} Discrete-phase precoding is commonly solved with trellis-based discrete-phase search \cite{kazemi2017tcomm}, alternating direction method of multipliers (ADMM) \cite{wang2018twc}, and penalty-based optimization of symbol-error probability \cite{shao2019tsp}.
    Quantized constant-envelope precoding with PSK and QAM signaling \cite{jedda2018twc, jacobsson2017quantized} and constant-envelope precoding with an adaptive receiver constellation \cite{zhang2016twc} also shape the received symbols directly.

    \item \textbf{Robust directional modulation.}
    Directional modulation designs channel excitations such that intended symbols are sent to legitimate receivers while reception elsewhere is impaired \cite{daly2009tap}.
    Few-bit phase control and imperfect channel information are considered in \cite{wei2021tcomm}, and errors in the measured user direction in \cite{shu2018sj}.

    \item \textbf{Annealing and decomposition methods.}
    Ising and quadratic unconstrained binary optimization (QUBO), simulated annealing (SA), and successive decomposition of TX array rows are performed for single objective optimization, such as scattering of reflective metasurfaces \cite{ross2022tap}, signal-to-noise ratio (SNR) \cite{krikidis2024lwc}, and spectral-efficiency \cite{yan2023twc}.

    \item \textbf{Iterative beamforming and beampattern synthesis.}
    Iterative methods such as successive refinement based on orthogonal matching pursuit, multi-point iterative gradient descent, and ADMM are used for hybrid digital and analog precoding for spatial multiplexing \cite{mirza2017lcomm}, beampattern synthesis of antenna arrays \cite{peng2022lcomm}, and sparse-array design \cite{huang2023tsp}.
\end{itemize}
Unlike prior approaches, which optimize spatial radiation patterns or suppress leakage to other directions, ODP synthesis fits each QAM symbol jointly over channels and subslots from finite-alphabet, oversampled phase states, and penalizes angular sensitivity.

In this work, ODP codeword synthesis and hardware calibration are formulated and cheaply solved.
The contributions are as follows:
\begin{enumerate}
    \item \textbf{Joint spatial and temporal codeword synthesis.}
    The ODP phased array is formulated to account for channel-dependent amplitude and phase errors, and the constellation-fitting objective is minimized without enumerating the full $(2^b)^{N \times T}$ phase configurations.

    \item \textbf{Tractable objective for angular robustness.}
    Under a second-order Taylor approximation of the steering response, the expected symbol-fitting error under Gaussian drift is a Hermitian positive-semidefinite (PSD) quadratic of rank at most three, which supports discrete search with inexpensive iterative updates. In ideal-hardware simulation, drift-aware synthesis reduces the EVM under drift by up to 1.54$\times$. On hardware, it reduces the measured EVM under drift by up to 1.18$\times$.

    \item \textbf{Solver comparison and random-start greedy.}
    Greedy, tiling, ADMM, and QUBO-SA are compared in accuracy, beam quality, and runtime across array sizes, temporal oversampling factors, and drift levels, and against exhaustive search for small arrays.
    Randomizing the channel order and forward-pass choices scatters cancellations across the aperture, increasing sidelobe suppression from 7.85 to 12.34~dB at $N=512$ and $T=1$ while EVM remains 4.51\%, at the same cost of greedy. It runs 13--16$\times$ faster than SA with cached response sums, with sidelobe suppression within 0.7~dB.
    
    \item \textbf{Cross-angle calibration with few pilots.}
    Effective phase-state gains shared across angles and angle-dependent channel responses are jointly estimated with alternating least squares (ALS). After full sweeps at the training angles, 16 pilot codewords at a new angle, excluded together with its neighbors from training, predict the other 240 measured responses of a 4-channel 22-nm FD-SOI array with mean errors of 5.46\% and 5.32\% for the 16- and 32-QAM datasets. This is up to 13.1$\times$ lower than an ideal-response model with a pilot-fitted common gain. Codebooks selected from these predictions reach measured EVMs of 7.98\% and 9.15\%.
\end{enumerate}

Sec.~\ref{sec:math} develops the array model, drift-aware synthesis objective, and calibration procedure. Sec.~\ref{sec:methodology} presents the synthesis algorithms. Sec.~\ref{sec:experiment} specifies the experimental protocols, and Sec.~\ref{sec:results} discusses the synthesis and calibration result. Sec.~\ref{sec:conclusion} concludes the paper. 

\section{Problem Formulation}
\label{sec:math}

\subsection{Channel Model and Array Factor}

In an $N$-channel TX, each channel consists of a constant envelope frequency quadrupler (FQ), PA, and a 2-bit vector-modulated PS. 
Channel $n$ selects one of 4-phase states $q[n]$ from ideal phasors $\mathcal{S}=\{1,\, j,\, -1,\, -j\}$ and excitation vector
\begin{equation}
    \mathbf{\omega} = \left[e ^ {j \frac{\pi}{2} q[n]}\right]^{N-1}_{n=0},
    \qquad 
    q[n] \in \mathcal{A} = \{0,1,2,3\}.
    \label{eq:steer}
\end{equation}
With channels at $\lambda/2$ spacing, the steering coefficient at observation angle measured from broadside $\theta$ is
\begin{equation}
    a(n;\theta) = e^{j \pi n \sin\theta}, \qquad n=0,\dots,N-1.
\end{equation}
The receiver (RX) sees the ideal normalized array factors over $N$ channels
\begin{equation}
    F(\theta; \mathbf{q})
    =
    \frac{1}{N} \sum_{n=0}^{N-1} \omega[n] a(n; \theta)
    =
    \frac{1}{N} \mathbf{\omega}^{\mathsf{T}}\mathbf{a}(\theta),
    \label{eq:af}
\end{equation}
where $\mathbf{q}=\big[q[n]\big]_{n=0}^{N-1}$ and $\mathbf{a}(\theta)=\left[a(n;\theta)\right]_{n=0}^{N-1}$.
For a desired QAM symbol $s$ at direction $\theta$, the TX selects $\mathbf{q}$ so that $F(\theta; \mathbf{q})$ best approximates $s$. The magnitude can vary from symbol to symbol because the channel outputs add with different relative phases at the receiver.

\begin{figure*}
    \centering
    \includegraphics[width=0.9\linewidth]{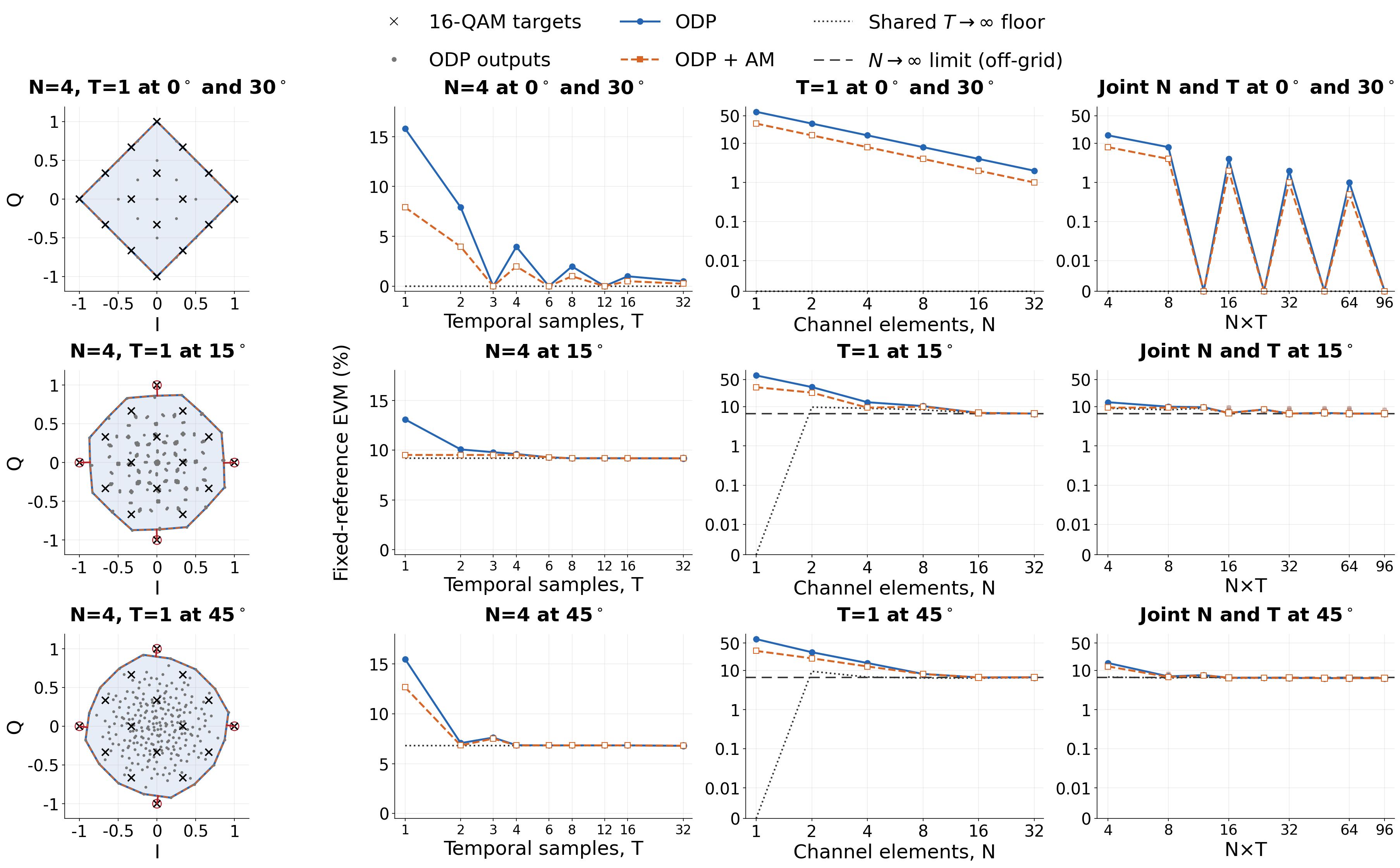}
    \caption{Amplitude control versus spatial and temporal resources at observation angles $\theta_s=0^\circ$, $30^\circ$, $15^\circ$, and $45^\circ$, with full-scale 16-QAM targets (corners at $|F|=1$ for every $N$) and half-wavelength spacing. 
    Left: at $N=4$ and $T=1$, 16-QAM targets (crosses), phase-only ODP outputs (gray dots), and the attainable hull of Eq.~\eqref{eq:hull}, which is identical without (blue solid) and with AM (orange dashed). Targets outside the hull are circled in red, but are still connected to their nearest attainable points.
    Others: exact optimal EVM (dots), ODP without (blue solid) and with AM (orange dashes) versus temporal samples $T$ at $N=4$, versus channel elements $N$ at $T=1$, and versus the product $NT$ (taking the best sampled factorization).}
    \label{fig:AM_vs_NT}
\end{figure*}
 
\subsection{Temporal Oversampling}
\label{subsec:odp}

With oversampling factor $T$, the TX divides one symbol interval into $T$ equal duration time subslots. In subslot $t$, channel $n$ selects a phase code $q_t[n]\in\mathcal A$ and applies the corresponding excitation
\begin{equation}
    \mathbf{\Omega}(\mathbf{Q}) = \Big[\underbrace{{\left[e ^ {j \frac{\pi}{2} q_t[n]}\right]}^{N-1}_{n=0}}_{\omega_t}\Big]^{T-1}_{t=0} = e ^ {j \frac{\pi}{2} \mathbf{Q}},
    \label{eq:code_new}
\end{equation}
where the phase states are
\begin{equation}
    \mathbf{Q}=\Big[\big[q_t[n]\big]_{n=0}^{N-1}\Big]_{t=0}^{T-1}, 
    \qquad 
    q_t[n] \in \mathcal{A} = \{0,1,2,3\}.
    \label{eq:Omega}
\end{equation}
The normalized array factor therefore is
\begin{equation}
    F(\theta; \mathbf{Q})
    =
    \frac{1}{N} \sum_{n=0}^{N-1} 
    \big(\frac{1}{T} \sum_{t=0}^{T-1} \omega_t[n] \big) a(n;\theta).
\end{equation}
Equivalently,
\begin{equation}
    F(\theta; \mathbf{Q})
    =
    \frac{1}{N} 
    \bar{\mathbf{\Omega}}^{\mathsf{T}}
    \mathbf{a}(\theta),
    \label{eq:af_new}
\end{equation}
where $\bar{\mathbf{\Omega}}=\frac{1}{T}\mathbf{\Omega}\mathbf{1}_T$ the effective excitation over the symbol interval. $\mathbf{1}_T$ is the $T \times 1$ column vector of ones.
Intuitively, QAM is not produced from individual channels, but from the coherent sum of many.

\subsection{Attainable Constellation}
\label{subsec:hull}

Fig.~\ref{fig:AM_vs_NT} compares the constellations synthesized by ODP with and without amplitude modulation (AM). The 16-QAM targets $F_m$ are rotated by $\frac{\pi}{4}$ and scaled by $\frac{1}{\sqrt{2}(\sqrt{M}-1)}$, so the four corner symbols lie on $\mathcal{S}$, the coherent peak $|F|=1$, for every $N$ and $T$. 

With or without AM, the symbol average of each channel lies in the diamond whose vertices are $\mathcal{S}$, and steering rotates the diamond of channel-$n$ by $\varphi_n=\pi n\sin\theta_s$. Every attainable array factor, therefore, lies in the average of the rotated diamonds, $\mathcal{H}_N(\theta_s)$.
A target $F$ lies in $\mathcal{H}_N(\theta_s)$ if and only if
\begin{equation}
    \small
    \operatorname{Re}(Fe^{-j\psi})
    \leq
    \underbrace{\frac{1}{N}\sum_{n=0}^{N-1}\max\big(|\cos(\psi-\varphi_n)|,\,|\sin(\psi-\varphi_n)|\big)}_{\eta_N(\psi)}
    \label{eq:hull}
\end{equation}
in every direction $\psi$ \cite{schneider1993convex}.

Specifically, to reach far in a direction $\psi$, every channel pushes as far as its diamond allows in that direction, which ranges from $\tfrac{1}{\sqrt{2}}$, when an edge faces $\psi$, to 1, when a corner points along $\psi$. $\eta_N(\psi)$ is the average of these reaches, and it equals 1 along the corner targets only at \textit{on-grid angles}, such as 0$^\circ$ and 30$^\circ$, where all diamonds share one orientation; other angles, such as 15$^\circ$ and 45$^\circ$, are \textit{off-grid}.

At on-grid angles, every target lies inside $\mathcal{H}_N$, and the targets are exactly attainable when $NT$ is a multiple of 3 \cite{kim2026jmw}. At off-grid angles, the corner targets lie outside $\mathcal{H}_N$, and their distance to the hull sets an EVM floor that no codeword search, oversampling factor, or AM can lower.

Nevertheless, AM reaches a given accuracy inside $\mathcal{H}_N$ with fewer channels or subslots. 
That is because AM permits an independent amplitude $\alpha_{n,t}\in[0,1]$ in each channel and subslot, whereas phase-only ODP fixes $\alpha_{n,t}=1$, which is compensated through additional subslots. 

At fixed $N$, increasing $T$ only fills $\mathcal{H}_N$ more densely, leaving the hull unchanged.
As $N$ grows at an off-grid angle, the diamond turns into an evenly spread disk of radius $r_\infty$. 
For 45$^\circ$, this radius is \cite{kuipers1974uniform}
\begin{equation}
    r_\infty(45^\circ)
    =\frac{2}{\pi}\int_{-\pi/4}^{\pi/4}\cos u\,du
    =\frac{2\sqrt{2}}{\pi}
    \approx 0.9.
\end{equation}
The four corner targets lie outside this disk, creating an EVM floor of
\begin{equation}
    100(1-r_\infty(45^\circ))\sqrt{\frac{4}{\sum_m|F_m|^2}}
    \approx
    6.69\%,
\end{equation}
as shown in Fig.~\ref{fig:AM_vs_NT}. 

\subsection{Channel and Phase-Shifter Non-idealities}

In practice, the channel response depends on both the observation angle and the phase state. 
Non-idealities of each PS give every phase state its own complex gain $\gamma(n;q)$. They arise before the signal is radiated, so they are the same at every angle. 
Everything after the PS, i.e., the fixed gain and phase of the channel's feed, the element's radiation toward $\theta$, and the propagation path to the receiver, is collected in the channel gain $e(n;\theta)$, which changes with the observation angle.

The array response is approximated by additive per-channel contributions. 
The effective steering vector is parameterized relative to the ideal steering vector as
\begin{equation}
    \small
    \mathbf{a}_e(\theta) = \mathbf{E}(\theta)\,\mathbf{a}(\theta),
    \qquad
    \mathbf{E}(\theta) = \operatorname{diag}\!\big(e_0(\theta),\dots,e_{N-1}(\theta)\big).
    \label{eq:ae}
\end{equation}

For codeword $\mathbf Q$, channel $n$ in subslot $t$ selects the gain $\gamma(n;q_t[n])$, giving
\begin{equation}
    \bar{\Omega}_{\Gamma}(\mathbf{Q}) = 
    \bigl[\frac{1}{T}\sum_{t=0}^{T-1} \underbrace{\gamma(n;q_t[n])\,
      e^{\,j\frac{\pi}{2}q_t[n]}}_{\omega_{t,\gamma}[n]}\bigr]^{N-1}_{n=0}.
    \label{eq:effcode1}
\end{equation}
To vectorize this, the gains are tabulated as $\boldsymbol{\Gamma}=\big[\gamma(n;q)\big]\in\mathbb{C}^{N\times4}$, with one row per channel and one column per phase state. The entries selected by $\mathbf{Q}$ are collected in $\boldsymbol{\Gamma}_{\mathbf{Q}} = \Big[\big[\gamma(n;q_t[n])\big]_{n=0}^{N-1}\Big]_{t=0}^{T-1}$, so
\begin{equation}
    \bar{\Omega}_{\Gamma}(\mathbf{Q}) = 
    \frac{1}{T}
    \underbrace{\left(\boldsymbol{\Gamma}_{\mathbf{Q}}\odot e^{j\frac{\pi}{2}\mathbf {Q}}\right)}_{\mathbf{\Omega}_{\Gamma}}\mathbf{1}_T
    = \frac{1}{T}\mathbf{\Omega}_{\Gamma} \mathbf{1}_T.
\end{equation}
The phase imbalance only carries how phase states differ from one another, but the channel gain is carried entirely by $\mathbf{E}$. To show this, the phase imbalance is normalized such that
\begin{equation}
    \frac{1}{4}\sum_{q\in\mathcal{A}} \gamma(n; q) = 1.
    \label{eq:gammanorm}
\end{equation}

\subsection{Pointing-Angle Drift}

Due to coarse steering or platform motion, the observation angle in Eq.~\eqref{eq:ae} is perturbed by a zero-mean Gaussian noise with $\mathbb{E}[\Delta\theta^2]=\sigma_\theta^2$ and $\mathbb{E}[\Delta\theta^4]=3\sigma_\theta^4$. 

Pointing drift is referenced to the array centroid, $\bar n=\tfrac{(N-1)}{2}$ for the element indices $n=0,\ldots,N-1$. The phase reference is chosen to preserve the nominal steering vector at $\theta_s$, giving
\begin{equation}
    \mathbf{a}(\theta_s+\Delta\theta)=\Big[e^{j\pi\left[(n-\bar n)\sin(\theta_s+\Delta\theta)+\bar n\sin\theta_s\right]}\Big]_{n=0}^{N-1},
    \label{eq:adrift}
\end{equation}
which reduces to $[e^{j\pi n\sin\theta_s}]_{n=0}^{N-1}$ at $\Delta\theta=0$. This convention removes the common phase term associated with referencing the array to element 0.

The Taylor expansion of the steering coefficient leads to the non-ideal array factor
\begin{equation}
    \small
    F(\theta_s + \Delta\theta) \approx \frac{1}{N} \bar{\Omega}_{\Gamma}^{\mathsf{T}}
    \left( \mathbf{a}_e(\theta_s) + \Delta\theta\,
    \frac{\partial\mathbf{a}_e}{\partial\theta}\bigg|_{\theta_s}
    + \frac{\Delta\theta^2}{2}\,
    \frac{\partial^2\mathbf{a}_e}{\partial\theta^2}\bigg|_{\theta_s} \right).
    \label{eq:perturb}
\end{equation}

\subsubsection{The Non-ideal Array Factor}

Over the drift range, the channel responses are held at their calibrated value $\mathbf{E}(\theta_s)$, and only the steering vector is expanded. The derivatives in Eq.~\eqref{eq:perturb}, or the drift sensitivity terms, are then
\begin{equation}
    \small
    \mathbf{d}_e(\theta_s) = \frac{\partial\mathbf{a}_e}{\partial\theta}\bigg|_{\theta_s}
    = \mathbf{E}(\theta_s)\big(\mathbf{a}(\theta_s)\odot\mathbf{c}\big),
    \label{eq:de}
\end{equation}
\begin{equation}
    \small
    \mathbf{h}_e(\theta_s) = \frac{\partial^2\mathbf{a}_e}{\partial\theta^2}\bigg|_{\theta_s} = \mathbf{E}\big(\mathbf{a}\odot(\mathbf{c}\odot\mathbf{c}+\mathbf{c}')\big),
    \label{eq:he}
\end{equation}
where 
\begin{equation}
    \small
    \mathbf{c}=\big[j\pi(n-\bar n)\cos\theta_s\big]_{n=0}^{N-1} \quad
    \mathbf{c}'=\big[-j\pi(n-\bar n)\sin\theta_s\big]_{n=0}^{N-1}.
    \label{eq:cs}
\end{equation}

Substituting Eq.~\eqref{eq:de} and Eq.~\eqref{eq:he} into Eq.~\eqref{eq:perturb} gives
\begin{equation}
    F(\theta_s+\Delta\theta) \approx
    \underbrace{\frac{1}{N} \bar{\mathbf{\Omega}}_{\Gamma}^{\mathsf{T}} \mathbf{a}_e(\theta_s)}_{\text{constellation point}}
    + \underbrace{\frac{\Delta\theta}{N} \bar{\mathbf{\Omega}}_{\Gamma}^{\mathsf{T}} \mathbf{d}_e(\theta_s)
    + \frac{\Delta\theta^2}{2N}\bar{\mathbf{\Omega}}_{\Gamma}^{\mathsf{T}}\mathbf{h}_e
    }_{\text{angle-drift error}}.
    \label{eq:Fnonideal}
\end{equation}
The steering sensitivity in $\mathbf{c}$ grows with the distance $n-\bar n$ from the array centroid, whereas the channel responses vary with the element pattern independently of $N$, so the steering term dominates as $N$ grows.

\subsection{Constellation Synthesis}

Following Eq.~\eqref{eq:code_new}, each RX only observes the subslot average, so two codewords are indistinguishable unless their averages differ. 
Let $D$ denote the number of distinct averages per channel.
Per channel, the average depends only on the multiset of the $T$ selected states, of which there are $\binom{T+3}{3}$. 
With non-ideal $\gamma(n;q)$, the multiset averages are generically distinct, so $D\le\binom{T+3}{3}$, with equality for generic gains.
In the ideal case, $\gamma(n;q)=1$, and phasors of $\mathcal{S}$ can cancel (e.g., $\{1,-1\}$ and $\{j,-j\}$ both average to 0), so the averages collapse onto the lattice points of the diamond with vertices $\mathcal{S}$ introduced in Subsec.~\ref{subsec:hull}, and $D=(T+1)^2$.
The synthesis algorithms search for one codeword per target symbol using the per-channel sets and the objective below, and enumerate all $D^N$ joint codewords only when a single tile spans the array.

For each target symbol $F$, let $\hat{F}_2$ denote the second-order Taylor approximation on the right-hand side of Eq.~\eqref{eq:Fnonideal}. The robust codeword-selection objective is the expected error of this approximation,
\begin{align}
    \small
    &J_F(\bar{\mathbf{\Omega}}_\Gamma)
    = \mathbb{E}\big[|F-\hat{F}_2|^2\big] \nonumber \\
    &= \Big|F-\mathbb{E}[\hat{F}_2]\Big|^2 + \operatorname{Var}(\hat{F}_2) \nonumber \\
    &= \underbrace{\Big| F - \tfrac{1}{N}\bar{\mathbf{\Omega}}_\Gamma^{\mathsf{T}}\,\tilde{\mathbf{a}}_e\Big|^{2}}_{\text{constellation fitting}} 
    + \underbrace{\frac{\sigma_\theta^{2}}{N^{2}}\Big|\bar{\mathbf{\Omega}}_\Gamma^{\mathsf{T}}\,\mathbf{d}_e\Big|^{2} + \frac{\sigma_\theta^4}{2N^2}\Big|\bar{\mathbf{\Omega}}_\Gamma^{\mathsf T}\mathbf{h}_e\Big|^2}_{\text{beam sensitivity}},
    \label{eq:robobj}
\end{align}
where $\tilde{\mathbf{a}}_e=\mathbf{a}_e+\frac{\sigma_\theta^2}{2}\mathbf{h}_e$.
$\tilde{\mathbf{a}}_e$, $\mathbf{d}_e$, and $\mathbf{h}_e$ are independent of the codeword, so expanding \eqref{eq:robobj} gives
\begin{align}
    \small
    J_F(\bar{\mathbf{\Omega}}_\Gamma) 
    &= \frac{1}{N^{2}}\,\bar{\mathbf{\Omega}}_\Gamma^\mathsf{H}\underbrace{\big(
    \tilde{\mathbf a}_e^{*}\tilde{\mathbf a}_e^{\mathsf T}+\sigma_\theta^2\,\mathbf d_e^{*}\mathbf d_e^{\mathsf T}+\tfrac{\sigma_\theta^4}{2}\,\mathbf h_e^{*}\mathbf h_e^{\mathsf T}
    \big)}_{\mathbf{R}}
    \bar{\mathbf{\Omega}}_\Gamma \nonumber \\
    &- \frac{2}{N}\operatorname{Re}\!\big(F^{*}\,\bar{\mathbf{\Omega}}_\Gamma^{\mathsf{T}}\tilde{\mathbf{a}}_e\big) + |F|^2,
    \label{eq:quadform}
\end{align}
where $\mathbf{R}$ is the sum of three outer products and thereby Hermitian PSD, and rank of at most 3. 
Here, the low rank keeps each iterative update inexpensive.
The constant $|F|^2$ does not change the minimizer.

$J_F$ is minimized once per symbol, and the resulting $M$ codewords form the codebook at $\theta_s$,
\begin{align}
    \mathbf{Q}^\star &\in \arg\min_{\mathbf{Q} \in \mathcal{A}^{N\times T}}J_F\big(\bar{\mathbf{\Omega}}_\Gamma(\mathbf{Q})\big).
    \label{eq:Qstar}
\end{align}

\subsection{Constellation Calibration}
\label{sec:estimation}

Parameters $n$ and $t$ describe what happens inside one codeword. Calibration measures many different codewords at several angles.
In other words, calibration assumes data is collected in a fixed temporal slot $T=1$, but over $S$ sessions, indexed by $s=0,\dots,S-1$, each at steering angle $\theta_s$ and containing codewords indexed by $k=0,\dots,K_s-1$.
In session $s$, codewords $\{\mathbf{Q}^{(s)}(k)\}_{k=0}^{K_s-1}$ are drawn.
For each, there is a measured normalized array factor $F(\theta_s;\mathbf{Q}^{(s)}(k))$, rendering the following objective,
\begin{equation}
    \small
    \min_{\boldsymbol{\Gamma},\,\{\mathbf{a}_e^{(s)}\}_{s=0}^{S-1}}
    \sum_{s=0}^{S-1}\sum_{k=0}^{K_s-1}
    \Bigl|
    F(\theta_s;\mathbf{Q}^{(s)}(k))
    - \tfrac{1}{N}\,
    \bar{\mathbf{\Omega}}_{\Gamma}^{\mathsf{T}}(\mathbf{Q}^{(s)}(k))\,
    \mathbf{a}_e^{(s)}
    \Bigr|^2.
    \label{eq:calibobj}
\end{equation}
to be solved. Eq.~\eqref{eq:calibobj} is bilinear in $\boldsymbol{\Gamma}$ and $\mathbf{a}_e^{(s)}$, motivating usage of alternating least squares (ALS) \cite{kolda2009tensor}.
Starting from an initial estimate of $\boldsymbol{\Gamma}$ and $\mathbf a_e^{(s)}$, the following updates are repeated:
\begin{enumerate}[leftmargin=*]
    \item
    Take $\mathbf{a}_e^{(s)}$ and update $\boldsymbol{\Gamma}$ using
    \begin{equation}\hspace{-10pt}
        \small
        \boldsymbol{\Gamma}^{\star}
        \leftarrow
        \arg\min_{\boldsymbol{\Gamma}}
        \sum_{s=0}^{S-1}\sum_{k=0}^{K_s-1}
        |F(\theta_s;\mathbf{Q}^{(s)}(k))
        - \tfrac{1}{N}\,
        \bar{\mathbf{\Omega}}_{\Gamma}^{\mathsf{T}}(\mathbf{Q}^{(s)}(k))\,
        \mathbf{a}_e^{(s)}
        |^2.
        \label{eq:gammaupdate}
    \end{equation}

    \item 
    Normalize each row of $\boldsymbol{\Gamma}$ to unit Euclidean norm for numerical stability:
    \begin{equation}
        \small
        \gamma(n;q)\leftarrow
        \frac{\gamma(n;q)}
        {\sqrt{\sum_{r\in\mathcal A}|\gamma(n;r)|^2}},
        \qquad q\in\mathcal A.
    \end{equation}

    \item
    Take $\boldsymbol{\Gamma}$ and update each $\mathbf a_e^{(s)}$ using
    \begin{equation}
        \small
        \mathbf{a}_e^{(s)\star}
        \leftarrow
        \arg\min_{\mathbf{a}_e^{(s)}}
        \sum_{k=0}^{K_s-1}
        |
        F(\theta_s;\mathbf{Q}^{(s)}(k))
        - \tfrac{1}{N}\,
        \bar{\mathbf{\Omega}}_{\Gamma}^{\mathsf{T}}(\mathbf{Q}^{(s)}(k))\,
        \mathbf{a}_e^{(s)}
        |^2.
        \label{eq:aeupdate}
    \end{equation}
\end{enumerate}
Repeat until convergence, and after the final iteration, impose Eq.~\eqref{eq:gammanorm}
by defining
\begin{equation}
    \mu_n=\frac14\sum_{q\in\mathcal A}\gamma(n;q),
\end{equation}
and for nonzero row means, rescaling
\begin{equation}
    \gamma(n;q)\leftarrow\frac{\gamma(n;q)}{\mu_n},
    \qquad
    a_e^{(s)}[n]\leftarrow \mu_n a_e^{(s)}[n]
\end{equation}
for every channel, phase state, and session.

\begin{figure*}
\centering
    \includegraphics[width=\linewidth]{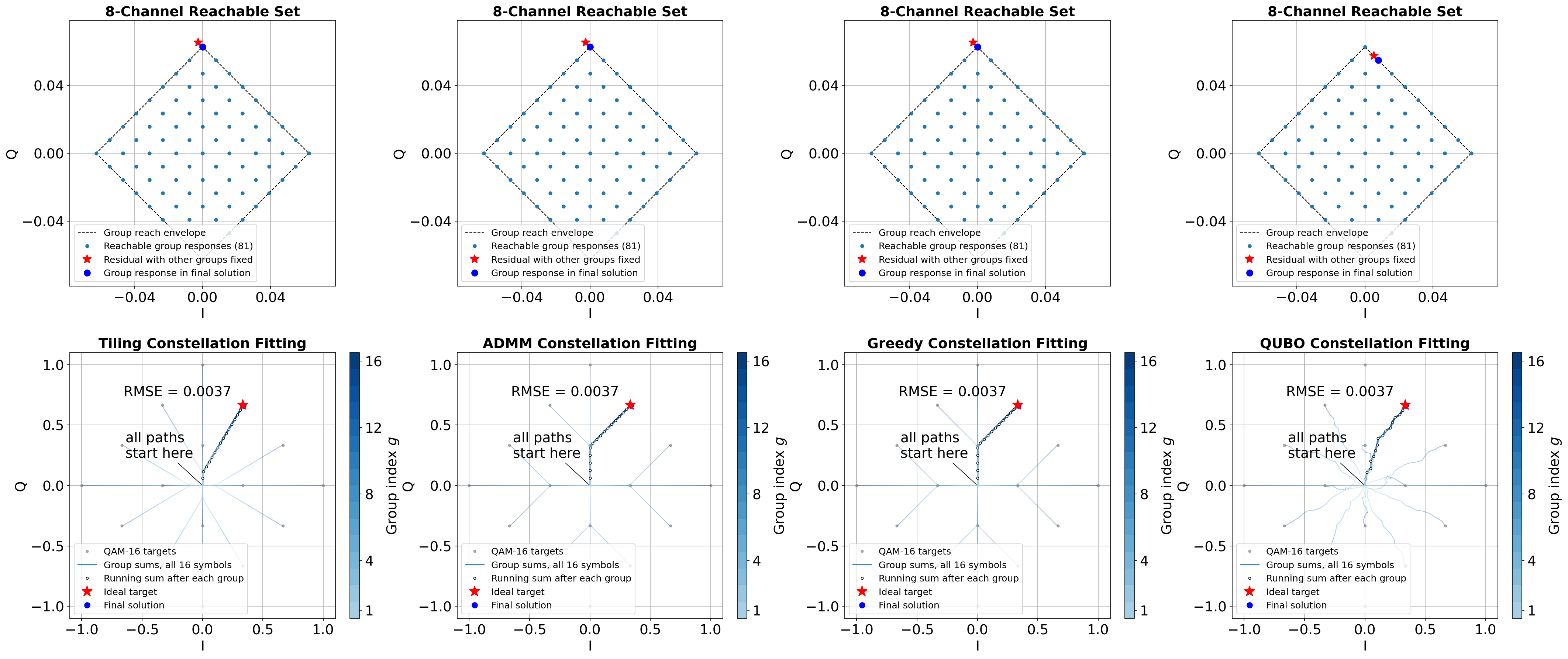}
    \caption{Tiling, ADMM, Greedy, and QUBO (left to right) for 16-QAM with $N=128$ and $\theta_s=0^\circ$. Top: eight-channel reachable sets. Bottom: constellation fitting through cumulative group sums. 
    Tiling optimizes small groups of channels together. 
    ADMM alternates between continuous fitting and snapping to allowed phase settings. 
    Greedy adjusts one channel at a time to reduce fitting error. 
    QUBO encodes phase choices as binary variables and uses simulated annealing to search for a good solution.}
    \label{fig:methodology}
\end{figure*}

\section{Methodology}
\label{sec:methodology}

Baseline and randomized greedy, tiling, ADMM, and QUBO-SA methods are presented in Algs.~\ref{alg:greedy}--~\ref{alg:qubo}, and their update structure is visualized in Fig.~\ref{fig:methodology}. 

All methods minimize the same per-symbol objective $J_F(\bar{\mathbf{\Omega}}_\Gamma)$ of Eq.~\eqref{eq:quadform}.
For channel-$n$, this is done by selecting one entry from its $D[n]$ distinct effective excitations and creating $\bar{\mathcal{S}}_\Gamma[n]=\{\bar{\Omega}^{(d)}_{\Gamma}[n]\}_{d=0}^{D[n]-1}$.
Each entry retains a representative subslot pattern, i.e., row $n$ of $\mathbf{Q}$, so the selected entries map directly to a codeword.
For each target $F$, the goal is solving Eq.~\eqref{eq:Qstar} to find the codeword whose array factor best approximates $F$. Repeating this for the $M$ targets closest to $M$-QAM forms the codebook. 
All algorithms are repeated for at most $N_\text{it}$ times.

For greedy and tiling refinement, in initial sweeps, objective-minimizing candidates are accepted, and ties are resolved by selecting the first candidate in a fixed ordering. A sweep with no changes terminates the search. If the initial sweep budget is exhausted, strict-decrease sweeps continue until no improvement exceeds a numerical tolerance or a second pass cap is reached. 

\subsection{Greedy}

\begin{algorithm}[t]
\caption{Greedy initialization and coordinate refinement}
\label{alg:greedy}
\begin{algorithmic}[1]
\Require $F$, $\bar{\mathcal{S}}_\Gamma[n]$, $J_F$, $\tau$, sweep caps and tolerance
\State $\pi\gets$ index order (baseline) or a random permutation (randomized)
\Function{ForwardGreedy}{$F$}
    \State $\bar{\mathbf{\Omega}}_\Gamma\gets\mathbf{0}$
    \For{$n=\pi_0,\ldots,\pi_{N-1}$}
        \State Take $\bar{\Omega}_\Gamma[n]$ from Eq.~\eqref{eq:greedy} if $\tau=0$; otherwise draw it using Eq.~\eqref{eq:softmax}
    \EndFor
    \State \Return $\bar{\mathbf{\Omega}}_\Gamma$
\EndFunction
\State $\bar{\mathbf{\Omega}}_\Gamma\gets\Call{ForwardGreedy}{F}$
\State Sweep channels in order $\pi$ using Eq.~\eqref{eq:greedy} with all others fixed
\State Apply the initial/strict refinement stopping rules
\State \Return the $\mathbf{Q}$ corresponding to $\bar{\mathbf{\Omega}}^\star_{\Gamma}$
\end{algorithmic}
\end{algorithm}

Greedy selection is a standard low-cost alternative to convex relaxation in combinatorial selection \cite{joshi2009convex}. The forward pass builds a feasible excitation vector one channel at a time, treating unassigned contributions as zero. For $\bar{\mathbf{\Omega}}_\Gamma[n\leftarrow v]$ denoting $\bar{\mathbf{\Omega}}_\Gamma$ with entry $n$ replaced by $v$, the update is
\begin{equation}
    \bar{\Omega}_\Gamma[n] \leftarrow
    \arg\min_{v\in\bar{\mathcal{S}}_\Gamma[n]}J_F(\bar{\mathbf{\Omega}}_\Gamma[n\leftarrow v]).
    \label{eq:greedy}
\end{equation}
Visiting the channels in index order makes the forward pass cancel residuals with adjacent channel pairs, which radiate as strong secondary lobes. The channels are therefore visited in a random order $\pi$, drawn once per symbol, and the forward pass draws candidate $v$ with probability \cite{feo1995greedy}
\begin{equation}
    p(v)\propto\exp\!\Big(-\frac{\Delta J_v}{\tau\,\overline{\Delta J}}\Big),
    \label{eq:softmax}
\end{equation}
where $\Delta J_v$ is the cost increase of $v$ over the best candidate of Eq.~\eqref{eq:greedy} and $\overline{\Delta J}$ is its mean over $\bar{\mathcal{S}}_\Gamma[n]$.
For $\tau=0$, or when all candidate costs are zero, the minimizer of Eq.~\eqref{eq:greedy} is taken, with ties going to the first entry of $\bar{\mathcal{S}}_\Gamma[n]$.
The standalone greedy solver subsequently revisits channels in order $\pi$ with all other selections fixed, using Eq.~\eqref{eq:greedy}, so it stops when no single-channel change lowers $J_F$ beyond the tolerance or the sweep cap is reached. Maintaining the three response sums in Eq.~\eqref{eq:quadform} gives a cost of $\mathcal{O}(N_{\text{it}}MND)$ per codebook.

\subsection{Tiling}

Tiling updates $R$ contiguous channels jointly, allowing coordinated discrete changes that a single-channel update cannot make. The array is partitioned into $P=\tfrac{N}{R}$ tiles. For each tile $p$, its attainable entries of $\bar{\mathbf{\Omega}}_\Gamma$ are enumerated and their response triples for $\tilde{\mathbf{a}}_e$, $\mathbf{d}_e$, and $\mathbf{h}_e$ are stored as in Eq.~\eqref{eq:Fnonideal}
\begin{equation}
    \begin{bmatrix}a_p(i)\\d_p(i)\\h_p(i)\end{bmatrix}
    =\frac{1}{N}\sum_{n\in\text{tile}\ p}\bar{\Omega}_{\Gamma,p}(i)[n]
    \begin{bmatrix}\tilde a_e[n]\\d_e[n]\\h_e[n]\end{bmatrix}.
    \label{eq:tiles}
\end{equation}
The tables are shared across targets at a fixed angle and drift setting. With the indices $i_{p'}$ of the other tiles $p'\neq p$ fixed, the candidate cost is
\begin{align}
    \small
    &J_p(i) = \Big|F-\sum_{p'\ne p}a_{p'}(i_{p'})-a_p(i)\Big|^2+ \nonumber\\
    &\sigma_\theta^2\Big|\sum_{p'\ne p}d_{p'}(i_{p'})+d_p(i)\Big|^2
    +\frac{\sigma_\theta^4}{2}\Big|\sum_{p'\ne p}h_{p'}(i_{p'})+h_p(i)\Big|^2.
    \label{eq:cost}
\end{align}
A single tile spanning the array gives exhaustive per-symbol search. Smaller tiles trade joint-search scope for lower table cost. At the end of each iteration, code adds up the contributions of all the other tiles.
Table construction costs $\mathcal{O}(ND^R)$ and refinement costs $\mathcal{O}(N_{\text{it}}MP(D^R+P))$, where $D^R$ is the number of candidate configurations of an $R$-channel tile. 

\begin{algorithm}[t]
    \caption{Tile-table construction and block refinement}
    \label{alg:tiling}
    \begin{algorithmic}[1]
    \Require $F$, $\bar{\mathcal{S}}_\Gamma[n]$, $R$ dividing $N$, sweep caps and tolerance
    \State Partition channels into $P=\tfrac{N}{R}$ contiguous tiles
    \State Cache Eq.~\eqref{eq:tiles} for each tile
    \State Initialize each tile against target $\tfrac{F}{P}$, with other tile contributions zero
    \State Sweep tiles cyclically; for each tile $p$:
    \Statex \hspace{\algorithmicindent}select $i_p\in\arg\min_i J_p(i)$ using Eq.~\eqref{eq:cost}
    \State Apply the initial/strict refinement stopping rules
    \State \Return the $\mathbf{Q}$ corresponding to $\bar{\mathbf{\Omega}}^\star_{\Gamma}$
\end{algorithmic}
\end{algorithm}

\subsection{ADMM}

ADMM \cite{boyd2011admm} couples a continuous excitation vector $\mathbf{x}$ to a feasible vector $\bar{\mathbf{\Omega}}_\Gamma$ through a scaled dual vector $\mathbf{u}$. 
The continuous step jointly adjusts all channels, while projection enforces their discrete states \cite{takapoui2020simple}. The continuous update is
\begin{equation}
    \mathbf{x}^{i+1}
    =\left(\frac{\mathbf{R}}{N^2}+\rho\mathbf{I}\right)^{-1}
    \left(\frac{F}{N}\tilde{\mathbf{a}}_e^*
    +\rho(\bar{\mathbf{\Omega}}_\Gamma^i-\mathbf{u}^i)\right),
    \label{eq:ADMM1}
\end{equation}
followed by
\begin{align}
    \bar{\Omega}_\Gamma^{i+1}[n]&\in\arg\min_{v\in\bar{\mathcal{S}}_\Gamma[n]}|x^{i+1}[n]+u^i[n]-v|^2,
    \label{eq:admmproj}\\
    \mathbf{u}^{i+1}&=\mathbf{u}^i+\mathbf{x}^{i+1}-\bar{\mathbf{\Omega}}_\Gamma^{i+1}.
    \label{eq:admmdual}
\end{align}
Since $\mathbf{R}$ has rank at most three, the Woodbury identity \cite{woodbury1950inverting} reduces the inverse to a system of size at most 3$\times$3. Scaling $\rho$ with $\tfrac{1}{N}$ keeps the penalty comparable to the normalized quadratic term as the array grows. The resulting cost is $\mathcal{O}(N_{\text{it}}MND)$ per codebook. 

For the warm-start initialization, the index-order forward pass of Alg.~\ref{alg:greedy} supplies an inexpensive feasible fit before these joint updates. 

\begin{algorithm}[t]
\caption{ADMM with greedy initialization and best-iterate retention}
\label{alg:admm}
\begin{algorithmic}[1]
\Require $F$, $\bar{\mathcal{S}}_\Gamma[n]$, $J_F$, $\rho$, $\epsilon$, $N_{\text{it}}$
\State $\bar{\mathbf{\Omega}}_\Gamma^0\gets\Call{ForwardGreedy}{F}$ if warm-start, else $\mathbf{0}$
\State $\mathbf{u}^0\gets\mathbf{0}$
\State $\bar{\mathbf{\Omega}}^\star_{\Gamma}\gets\bar{\mathbf{\Omega}}_\Gamma^0$, $J^\star\gets J_F(\bar{\mathbf{\Omega}}_\Gamma^0)$ if warm-start, else $J^\star\gets+\infty$
\For{$i=0,\ldots,N_{\text{it}}-1$}
    \State Update $\mathbf{x}^{i+1}$ using Eq.~\eqref{eq:ADMM1}
    \State Project each channel using Eq.~\eqref{eq:admmproj}
    \State Update $\mathbf{u}^{i+1}$ using Eq.~\eqref{eq:admmdual}
    \If{$J_F(\bar{\mathbf{\Omega}}_\Gamma^{i+1})<J^\star$}
        \State $\bar{\mathbf{\Omega}}^\star_{\Gamma}\gets\bar{\mathbf{\Omega}}_\Gamma^{i+1}$, $J^\star\gets J_F(\bar{\mathbf{\Omega}}_\Gamma^{i+1})$
    \EndIf
    \If{$\|\mathbf{x}^{i+1}-\bar{\mathbf{\Omega}}_\Gamma^{i+1}\|_2<\epsilon$ and $\rho\|\bar{\mathbf{\Omega}}_\Gamma^{i+1}-\bar{\mathbf{\Omega}}_\Gamma^i\|_2<\epsilon$}
        \State \textbf{break}
    \EndIf
\EndFor
\State \Return the $\mathbf{Q}$ corresponding to $\bar{\mathbf{\Omega}}^\star_{\Gamma}$
\end{algorithmic}
\end{algorithm}

\subsection{QUBO-SA}

QUBO-SA encodes the channel excitations in a binary vector $\mathbf{b}$ and minimizes $E(\mathbf{b})=J_F(\mathbf{z}(\mathbf{b}))+\sum_n\mathcal{P}[n]$, where $\mathcal{P}[n]$ penalizes invalid encodings in \cite{lucas2014ising}
\begin{equation}
    \small
    \mathcal{P}^{\text{one-hot}}[n]
    = \lambda_{\text{pen}}
    \left(\sum_{p=0}^{D[n]-1} b_{p}[n]-1\right)^2,
    \quad 
    b_{p}[n]\in\{0,1\}.
    \label{eq:QUBOpenalty}
\end{equation}
The 4 attainable excitations are represented by 2 selection bits and an auxiliary bit constrained to equal their product. For larger sets, one bit represents each attainable excitation, with exactly one bit active per channel. Substituting the mappings into Eq.~\eqref{eq:quadform} with the additional quadratic constraint penalties yields a quadratic binary objective. The implemented annealer updates entire channel blocks while preserving valid encodings.

Each restart uses a random feasible initialization and geometric cooling, which multiplies the annealing factor by $\tau_\ell$, as in
\begin{equation}
    \tau_0=\frac{\Delta\hat E_+}{\log2},\qquad
    \tau_\ell=\tau_0(10^{-3})^{\frac{\ell}{N_{\text{it}}}}.
    \label{eq:annealschedule}
\end{equation}
The initial temperature is set from the median positive energy change $\Delta\hat E_+$ of trial proposals at a separate random feasible state.

If no trial increases the energy, $\Delta\hat E_+=1$ is used. Annealing accepts a temporary increase to leave a local minimum, at the cost of many proposals and restarts \cite{kirkpatrick1983optimization}. With $b$-bits per channel, a conservative dense QUBO cost bound is $\mathcal{O}(N_{\text{r}}N_{\text{it}}M(Nb)^2)$ per codebook, and with cached sums $O(N_{\text{r}} N_{\text{it}} M N)$.

\begin{algorithm}[t]
\caption{QUBO-SA with feasible channel-block proposals}
\label{alg:qubo}
\begin{algorithmic}[1]
\Require $F$, $\bar{\mathcal{S}}_\Gamma[n]$, $N_{\text{it}}$, restart count $N_{\text r}$
\State Build $E(\mathbf b)$; build $\tau_0$
\State $E_{\text{best}}\gets+\infty$
\For{each restart}
    \State Draw a random feasible $\mathbf b$; update the best visited state
    \For{$\ell=0,\ldots,N_{\text{it}}-1$}
        \State Set $\tau_\ell$ using Eq.~\eqref{eq:annealschedule}
        \For{$N$ proposals}
            \State Draw a channel and different phase states for it uniformly
            \State Rewrite its binary block to obtain feasible $\mathbf b'$
            \State $\Delta E\gets E(\mathbf b')-E(\mathbf b)$
            \State Accept with probability $\min\{1,e^{-\tfrac{\Delta E}{\tau_\ell}}\}$
            \State Retain the lowest-energy feasible state visited
        \EndFor
    \EndFor
\EndFor
\State \Return the $\mathbf{Q}$ corresponding to $\bar{\mathbf{\Omega}}^\star_{\Gamma}$
\end{algorithmic}
\end{algorithm}

\section{Experimental Setting}
\label{sec:experiment}

\subsection{Simulated Synthesis}
\label{subsec:scenario}

Solvers of Sec.~\ref{sec:methodology} are compared on 16-QAM over $N\in\{2^2,2^3,\dots,2^9\}$ channels, $T\in\{1,4\}$ subslots, and nine steering angles $\theta_s\in\{-40^\circ,-30^\circ,\dots,40^\circ\}$.

\subsubsection{Solver Settings}

Table~\ref{tab:solver_settings} lists the solver parameters. 
The iteration and sweep caps were increased until EVM stopped improving.
The solver benchmark uses ideal hardware, $\mathbf{E}=\mathbf{I}$ and $\gamma(n;q)=1$, so $\mathbf{a}_e=\mathbf{a}(\theta)$.

\begin{table}[t]
    \centering
    \caption{Solver settings.}
    \label{tab:solver_settings}
    \footnotesize
    \begin{tabular}{@{}lll@{}}
        \toprule
        Method & Parameter & Value \\
        \midrule
        Greedy & Refinement sweeps & 5 \\
               & Baseline order, $\tau$ & index order, 0 \\
               & Randomized order, $\tau$ & random per symbol, 0.3 \\
        \midrule
        Tiling & $R$ & largest in $\{8,4,2,1\}$ such that \\
               & & $R\mid N$ and $D^R\leq4\times10^{6}$ \\
               & Initial sweeps & 4 \\
        \midrule
        Tiling, greedy & Strict-decrease sweeps & 1000 \\
                       & Decrease tolerance & $10^{-12}(J_F+|F|^2)$ \\
        \midrule
        ADMM & $\rho$ & $\tfrac{10^{-3}}{N}$ \\
             & $N_{\text{it}}$ & 1000 \\
             & $\epsilon$ & $10^{-6}$ \\
        \midrule
        QUBO-SA & $\lambda_{\text{pen}}$ & $\max(20,5N)$ \\
                & $N_{\text{r}}$ & 5 \\
                & $N_{\text{it}}$ & 500, $N$ proposals each \\
                & Trials for $\Delta\hat{E}_+$ & 256 \\
        \bottomrule
    \end{tabular}

    \vspace{2pt}
\end{table} 

\subsubsection{Pointing Drift}

To compare robustness across array sizes at a common relative pointing uncertainty, the drift is scaled to the beamwidth, $\sigma_\theta=\kappa\,\theta_{\text{3dB}}$ with $\kappa\in\{0,0.2\}$, where $\theta_{\text{3dB}}=\tfrac{1.77}{N\cos\theta_s}$~rad is the nominal half-power beamwidth of a uniformly excited half-wavelength linear array.
$\kappa=0.2$ would translate to a drift of $0.16^\circ$ for $N=128$ and $1.27^\circ$ for $N=16$ at broadside, which by the standard pointing-loss approximation, costs about 0.5~dB of gain \cite{balanis1982antenna}.

The codebook is designed with the Taylor model of Eq.~\eqref{eq:quadform} but evaluated with the exact array factor $\hat{F}_m(\Delta\theta)=\tfrac{1}{N}\bar{\mathbf{\Omega}}_\Gamma^{\mathsf{T}}(\mathbf{Q}_m)\,\mathbf{a}(\theta_s+\Delta\theta)$ of Eq.~\eqref{eq:adrift}.

\subsection{Performance Metrics}

Constellation accuracy over the $M$ targets is reported as
\begin{equation}
    \small
    \text{EVM}_{\text{drift}}\big|_{\kappa_\text{design}}
    =100\times\sqrt{\frac{\sum_{m=0}^{M-1}\mathbb{E}_{\Delta\theta}\big[|F_m-\hat{F}_m(\Delta\theta)|^2\big]}{\sum_{m=0}^{M-1}|F_m|^2}},
    \label{eq:driftevm}
\end{equation}
and the nominal EVM\textsubscript{ideal} is Eq.~\eqref{eq:driftevm} at $\Delta\theta=0$.
The expectation over $\Delta\theta$ is computed by three-point Gauss--Hermite quadrature \cite{golub1969calculation},
\begin{equation}
    \small
    \text{EVM}_{\text{drift}}=\sqrt{\tfrac{2}{3}\text{EVM}^2(\theta_s)+\tfrac{1}{6}\text{EVM}^2(\theta_-)+\tfrac{1}{6}\text{EVM}^2(\theta_+)},
    \label{eq:gh}
\end{equation}
where $\theta_\pm=\theta_s\pm\sqrt{3}\sigma_\theta$ and $\text{EVM}(\theta)$ is the EVM of the codebook observed at $\theta$.

In simulation, the reference is known, so no gain or phase is refitted. On hardware, measured responses have no absolute scale or phase, so the EVM at each angle removes one complex gain fitted to the $M$ responses scored there. Prediction error fits no gain.

The robustness gain is
\begin{equation}
    G=\frac{\mathrm{EVM}_\text{drift}\big|_{\kappa_\text{design}=0}}{\mathrm{EVM}_\text{drift}\big|_{\kappa_\text{design}=\kappa}},
    \label{eq:G}
\end{equation}
where both codebooks are evaluated under the same drift $\kappa$. $G>1$ indicates that drift-aware design reduces EVM\textsubscript{drift}.

Beam quality is evaluated from the mean magnitude pattern 
\begin{equation}
    A(\theta)=\frac{1}{M}\sum_{m=0}^{M-1} \big| F(\theta;\mathbf{Q}_m) \big|.
\end{equation}
Sidelobe suppression is 
\begin{equation}
    20\log_{10}\Big(\frac{A_{\text{pk}}}{A_{\text{sl}}}\Big),
\end{equation}
where $A_{\text{pk}}$ is the peak within $\theta_s\pm\frac{12}{N}$~rad and $A_{\text{sl}}$ is the largest value over the visible region outside the main lobe, which is bounded by the first minimum of $A(\theta)$ on each side of the peak.

The pointing error is
\begin{equation}
    \text{Pointing Error}=\theta_\text{peak}-\theta_s
\end{equation}

Runtime is the wall-clock time of a complete solver call for all $M$-symbols, including initialization and table construction, in a single Python process on an Intel Core i9-13900H CPU. 

Unless an angle group is specified, reported values are averaged over the steering angles, and figures additionally show the range over angles.

\subsection{Measurement Protocol}

The calibration model is evaluated on measurements of a 4-channel 22-nm FD-SOI transmit array \cite{kim2026jmw} at 120 GHz, with 10 measured angles in the 16-QAM dataset ($0^\circ$, $5^\circ$, $\pm10^\circ$, $\pm20^\circ$, $\pm30^\circ$, $\pm40^\circ$) and 13 in the 32-QAM dataset ($0^\circ$, $\pm5^\circ$, $\pm10^\circ$, $\pm15^\circ$, $\pm20^\circ$, $\pm30^\circ$, $\pm40^\circ$).
The datasets are separate measurement sessions. Because each angle contains all $4^4=256$ codewords at $T=1$, either can be used to synthesize any constellation.

\begin{figure}
    \centering
    \includegraphics[width=1\columnwidth]{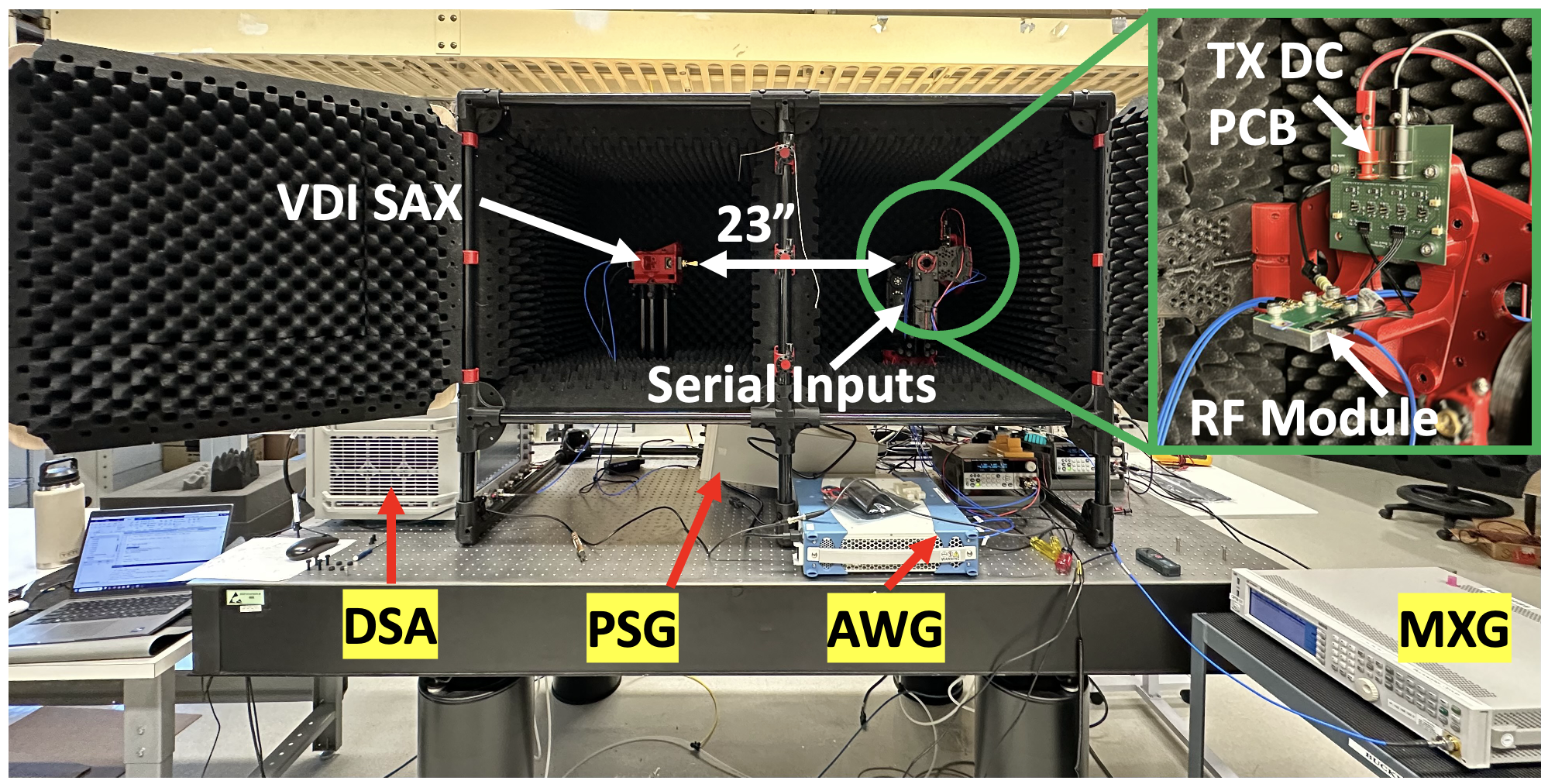}
    \caption{Photograph of the OTA measurement setup with the experimental 4-channel D-band transmit array (right) separated by 23 inches from the high-gain receive antenna (left).}
    \label{fig:MeasSetupPhoto}
\end{figure}

The over-the-air (OTA) measurement setup for the experiments is shown in Fig. \ref{fig:MeasSetupPhoto}. The 4-channel TX prototype is mounted on a Millibox GIM04 antenna positioner with serial data and clock signals generated with a Keysight M8195A AWG. The local oscillator (LO) signal is generated off chip using a Keysight MXG analog signal generator. In all measurements, the RX was placed 23 inches away from the TX. The RX side consists of a WR-6.5 horn antenna connected to a VDI WR6.5SAX-M extender with its intermediate frequency (IF) signal sampled by a Keysight DSAZ634A oscilloscope. Keysight vector signal analysis (VSA) software is used to resample the waveform from the digital signal analyzer (DSA), while MATLAB is used to control the VSA and post-process the data \cite{Tervo2026tmtt}.

\subsection{Calibration Protocol}
\label{subsec:calprotocol}

For each dataset, one angle is \textit{held-out} at a time, and the angles other than it and its measured neighbors are called \textit{in-sample}. The phase-state gains $\gamma(n;q)$ and the in-sample $\mathbf{a}_e$ are fitted to all 256 codewords at the in-sample angles as in Sec.~\ref{sec:estimation}. After, $\gamma(n;q)$ is fixed, and $\mathbf{a}_e$ at the held-out angle is estimated from $m_p$-\textit{pilot codewords}, i.e., known codewords whose measured responses are used for this estimate. 

\subsubsection{Prediction}

Prediction error is EVM\textsubscript{ideal} of Eq.~\eqref{eq:driftevm} with the sums taken over the remaining $256-m_p$ codewords, $F$ the measured and $\hat{F}$ the predicted responses, averaged over 25 random pilot draws per held-out angle. It measures agreement between predicted and measured responses.
The \textit{in-sample} reference fits the model to all 256 codewords at every angle, so its error is the in-sample fit error.

\subsubsection{Synthesis}

Since measured responses have no absolute scale, the target QAM is scaled to 0.60--1.00$\times$ the largest predicted magnitude and rotated by 0$^\circ$--89$^\circ$, keeping the pair with the lowest predicted EVM. With $N=4$ and $T=1$, all 256 codewords are evaluated, and instead of the per-symbol minimization of Eq.~\eqref{eq:Qstar}, each target is assigned a distinct codeword by exact linear assignment \cite{kuhn1955hungarian}, so the result depends only on the model, not on a solver.

To test synthesis, each model uses the same pilots to predict and synthesize the codebook at the held-out angle, and the selected codewords are scored with their measured responses there.
The \textit{in-sample} reference selects the codebook from the measured responses themselves.

\subsubsection{Drift-awareness}

To test drift-aware synthesis, the codebook designed at a held-out $\theta_s$ is also scored at its measured neighbors $\theta_{s\pm1}$, $h$ away on each side. Setting $\sigma_\theta=\tfrac{h}{\sqrt{3}}$, i.e., $0.11\lesssim \kappa \lesssim0.22$, places $\theta_\pm$ of Eq.~\eqref{eq:gh} on $\theta_{s\pm1}$.
Drift-aware synthesis uses $J_{F_m}(\bar{\mathbf{\Omega}}_\Gamma(\mathbf{Q}))$ of Eq.~\eqref{eq:robobj} at this $\sigma_\theta$ as the assignment cost of target $F_m$ and codeword $\mathbf{Q}$, and to choose the scale and rotation.

Only held-out angles with two equally spaced measured neighbors are tested, 6 of 10 (16-QAM) and 9 of 13 (32-QAM), e.g., $0^\circ$ in the 16-QAM dataset is skipped because its neighbors are $-10^\circ$ and $5^\circ$.

\section{Experimental Results}
\label{sec:results}

\subsection{Model Accuracy}

\begin{figure}
    \centering
    \includegraphics[width=\linewidth]{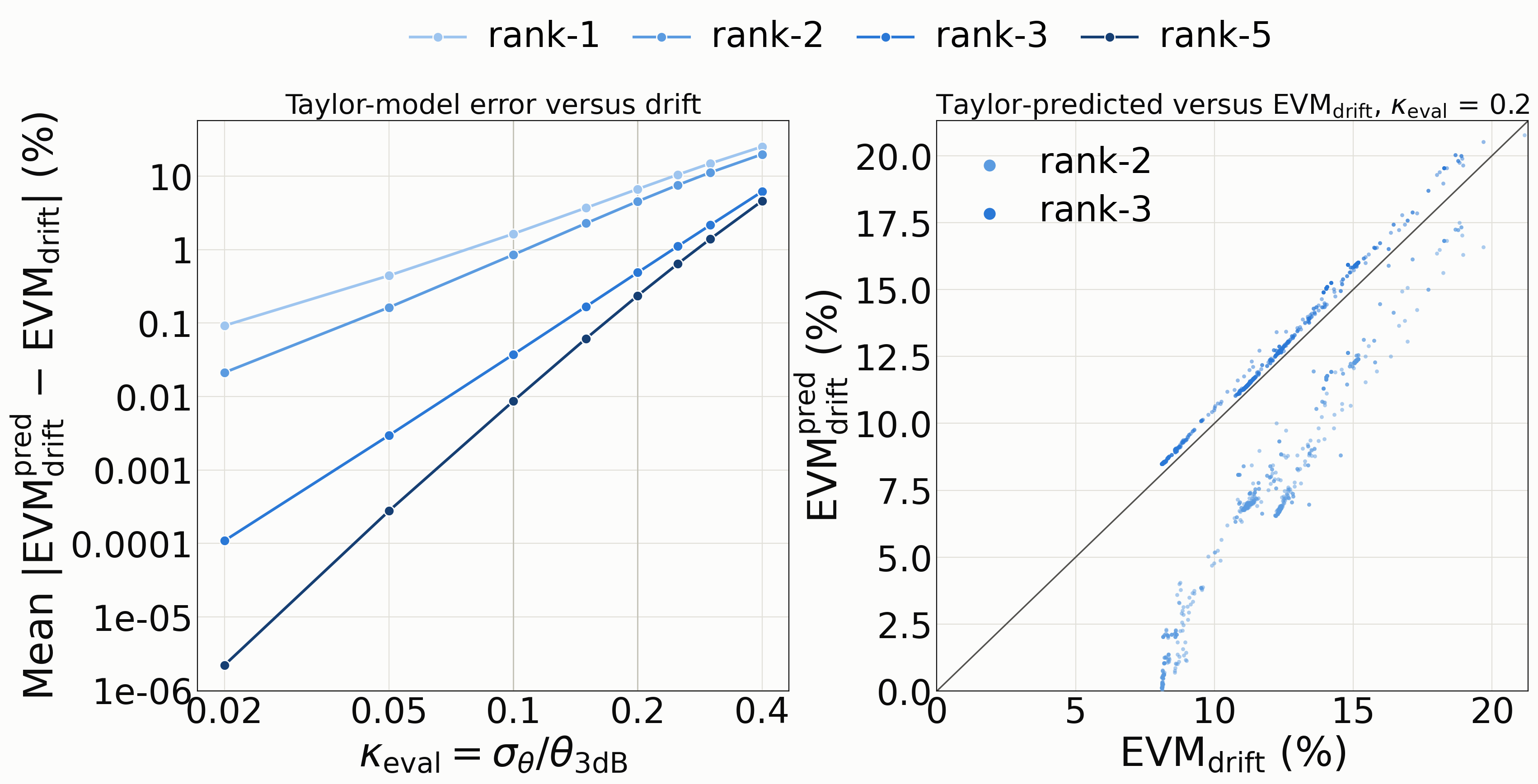}
    \caption{Taylor drift models versus EVM\textsubscript{drift} of 16-QAM codebooks.
    Left: mean absolute error versus $\kappa_\text{eval}$ for rank-1, -2, -3, and -5 models.
    Right: predicted versus EVM\textsubscript{drift} at $\kappa_\text{eval}=0.2$, one dot per codebook.}
    \label{fig:Taylor}
\end{figure}

The drift objective of Eq.~\eqref{eq:robobj} and the rank-3 $\mathbf{R}$ of Eq.~\eqref{eq:quadform} follow from truncating the Taylor expansion at the second order. 
Fig.~\ref{fig:Taylor} tests this truncation against EVM\textsubscript{drift} for 16-QAM codebooks of the five methods with settings of Subsec.~\ref{subsec:scenario}, each evaluated at $\kappa_\text{eval}\in\{0.02,0.05,0.1,0.15,0.2,0.25,0.3,0.4\}$.

At $\kappa_\text{eval}=0.2$, EVM\textsubscript{drift} averages 12.01\%. Ignoring drift (rank-1) predicts 5.35\%, and keeping only the first order (rank-2) 7.48\% instead, as both omit the curvature of the main lobe.
The second-order model predicts 12.50\%, and completing the expansion to the fourth order, predicts 12.25\%. The second-order model is therefore the lowest order that captures the main-lobe curvature, and it lands on the conservative side, overestimating EVM\textsubscript{drift} for every codebook.

\subsection{Solver Comparison}
\label{subsec:comparison}

\begin{figure}
    \includegraphics[width=\linewidth]{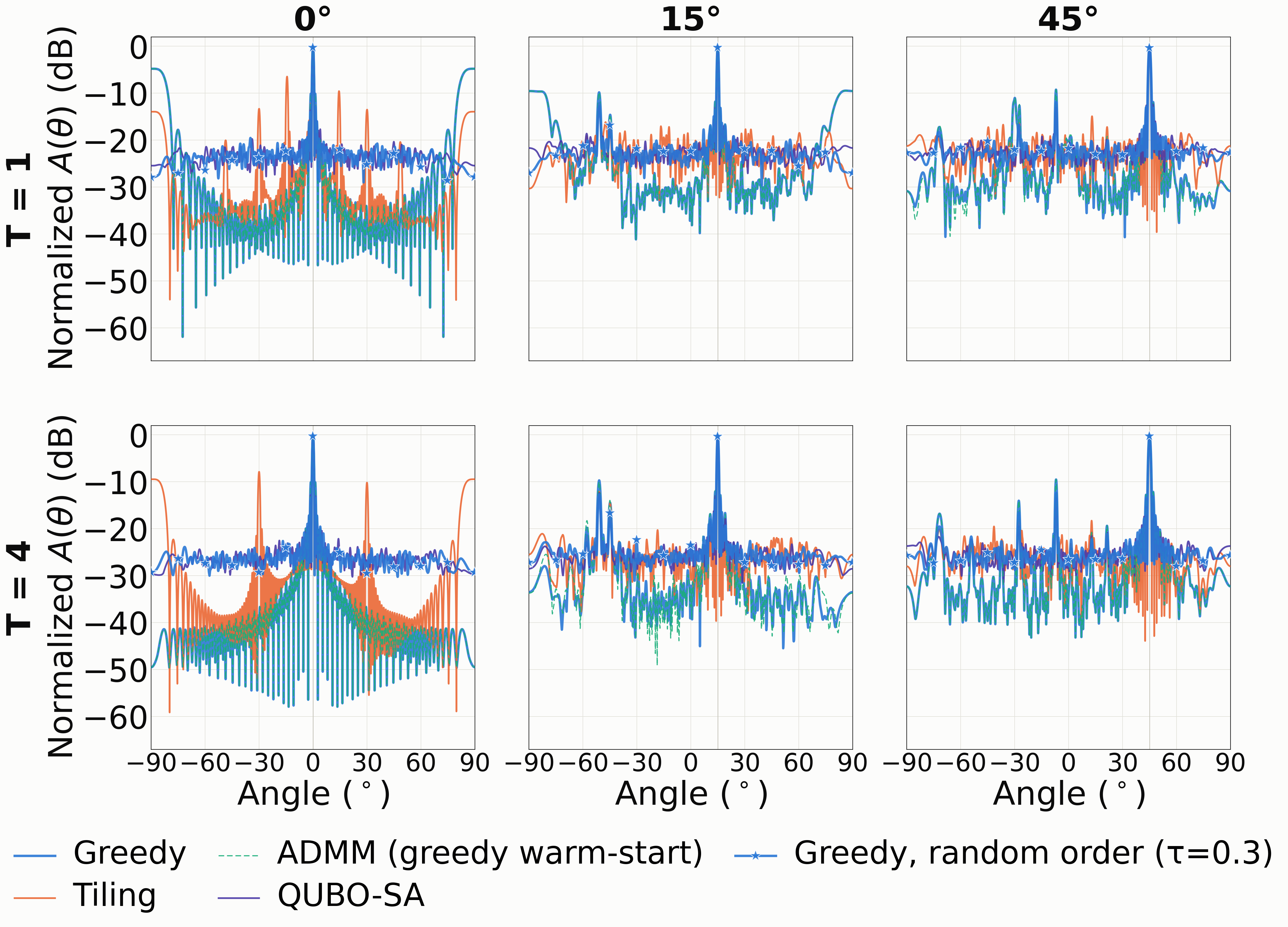}
    \caption{Mean magnitude pattern over the 16-QAM codewords, normalized to its peak, for $N=128$, $\kappa=0$, $T\in\{1,4\}$, and steering angles $0^\circ$, $15^\circ$, and $45^\circ$. ADMM uses a greedy warm start.}
    \label{fig:beampatterns}
\end{figure}

\begin{figure*}
    \includegraphics[width=\linewidth]{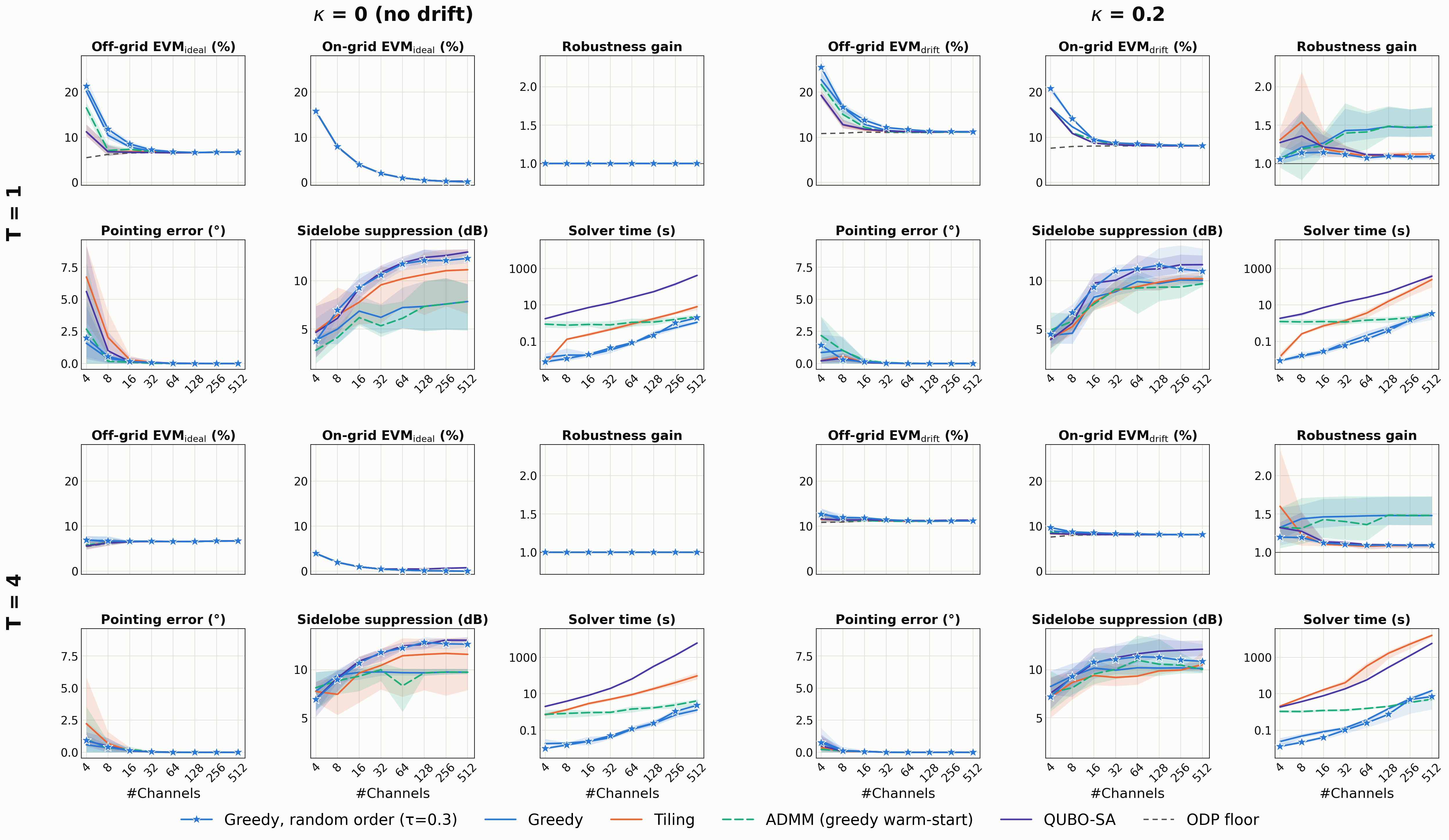}
    \caption{16-QAM synthesis over $N=4-512$-channels, $T\in\{1,4\}$, and $\kappa\in\{0,0.2\}$. Each block reports off-grid and on-grid EVM, robustness gain $G$ of Eq.~\eqref{eq:G}, absolute pointing error, sidelobe suppression, and solver runtime. Lines and bands show means and ranges over the relevant steering angles from $-40^\circ$ to $40^\circ$ in $10^\circ$ steps.
    The ODP floor is a lower bound obtained by relaxing the per-channel codebook to its convex hull \cite{joshi2009convex}.}
    \label{fig:comparison}
\end{figure*}

At $(N,T,\kappa)=(256,1,0)$, cold-started-ADMM step first assigns the same amplitude $\tfrac{|F|}{(1+N\rho)}\approx|F|$ to every channel, which unit-magnitude states at $T=1$ cannot represent. ADMM with greedy-warm-start lowers nominal EVM from 26.01\% to 4.55\%; therefore, it is chosen as the baseline ADMM instead.

At $(N,T,\kappa)=(512,1,0)$, sidelobe suppression is 7.9~dB for baseline greedy, 7.9~dB for warm-started-ADMM, 11.2~dB for tiling, and 13.0~dB for QUBO-SA. Baseline greedy and warm-started-ADMM favor adjacent opposite-phase cancellations, tiling repeats tile-level arrangements, and QUBO-SA's random starts spread cancellations across the aperture, producing a lower sidelobe floor, as illustrated in Fig.~\ref{fig:beampatterns}. 
Increasing $T$ to 4 lets channels reduce amplitude through subslot averaging, raising baseline greedy and warm-started-ADMM suppression from 7.9 to 9.7~dB. 

The local fits tend to center the pattern peak away from $\theta_s$ in $N\leq32$, but for $N\geq128$ the mean absolute pointing error of every method is below 0.01$^\circ$.

Greedy-based approaches (e.g., baseline greedy and greedy-warm-started-ADMM) concentrate the coherent contribution near one end, making it sensitive to the pointing drift. Tiling and random-start QUBO-SA, on the other hand, distribute it more evenly. 
Penalizing angular sensitivity reduces EVM\textsubscript{drift} the largest in tiling from 16.10\% to 10.43\% at $(N,T,\kappa)=(4,4,0.2)$, a reduction of 1.54$\times$.
Nevertheless, with drift-aware design, all five solvers reach nearly the same EVM\textsubscript{drift} for $N\geq128$, approximately 10\% at $\kappa=0.2$. This means all five solvers end up equally robust, and a large robustness gain does not make greedy or ADMM better at handling drift, but it means they had more sensitivity to remove. 
Drift-awareness also costs little when pointing is perfect, as at $N=512$, designing for $\kappa=0.2$ increases EVM\textsubscript{ideal} by at most 0.30 precentage points across all solvers and $T$.

As evident in Fig.~\ref{fig:comparison}, low EVM at the intended angle alone guarantees neither low sidelobes nor robustness to pointing drift. That is because Eq.~\eqref{eq:quadform} constrains the response only at $\theta_s$ and its first two derivatives \cite{er1983derivative, vorobyov2003robust}.
Nevertheless, the pointing drift depends on where the coherent and canceling channels lie along the aperture. 
Therefore, the power removed by cancellation is radiated at $\theta \neq \theta_s$. 
At $(N,T)=(4,1)$, however, few of the 256 codewords have both low nominal error and low angular sensitivity, and drift-aware tiling increases EVM\textsubscript{ideal} from 12.74\% to 17.14\%.

Tiling scans $(T+1)^{2R}$ candidates per tile, a runtime cost of $\sim3.9\times10^5$ at $T=4$, and dense QUBO-SA grows quadratically in the number of binary variables. 
At the array sizes studied here, runtime and computational complexity of ADMM differ, but they match for greedy.
Each ADMM iteration is a few vectorized operations whose runtime stays fixed, but in greedy, each channel is visited one at a time, the runtime of which grows linearly.
Nevertheless, QUBO-SA finds high-quality beams with low pointing error and high sidelobe suppression.

Random channel order alone provides 10.71~dB, and the softmax forward pass of Eq.~\eqref{eq:softmax} the remaining 1.6~dB. At $T=4$ and $N\leq8$, randomization does not help, and at $(4,4,0)$ it lowers suppression from 7.76 to 6.32~dB. 
Annealing scores each proposal in constant time once the three response sums are cached, so the dense QUBO matrix, not annealing itself, dominates QUBO-SA's runtime: at $(512,4,0)$ the dense implementation took $6.10\times10^3$~s. With cached sums, the same annealer takes 23.1~s on an idle machine and randomized greedy 1.47~s, 15.7$\times$ faster (13.3$\times$ at $T=1$). Randomized greedy reaches a lower EVM (4.48\% versus 4.75\%) and 0.40~dB lower suppression (12.66 versus 13.06~dB). Annealing with fewer sweeps trades EVM for sidelobes. At $(512,1,0)$, 50 sweeps (4.5~s) give 4.66\% and 13.03~dB, against 4.51\% and 12.34~dB for randomized greedy. Randomization therefore recovers most of the sidelobe advantage of annealing at greedy cost.

At $(N,T,\kappa)=(512,1,0)$, sidelobe suppression increases from 7.85 in baseline greedy to 12.34~dB in the randomized greedy, while EVM remains 4.51\%. 
At $N=4$, a single tile spans the array, so tiling equals exhaustive per-symbol search, whereas baseline and randomized greedy remain 0.9--6.1 precentage points above it in nominal EVM.

\subsection{Hardware Calibration}

\begin{figure*}
    \centering
    \includegraphics[width=\linewidth]{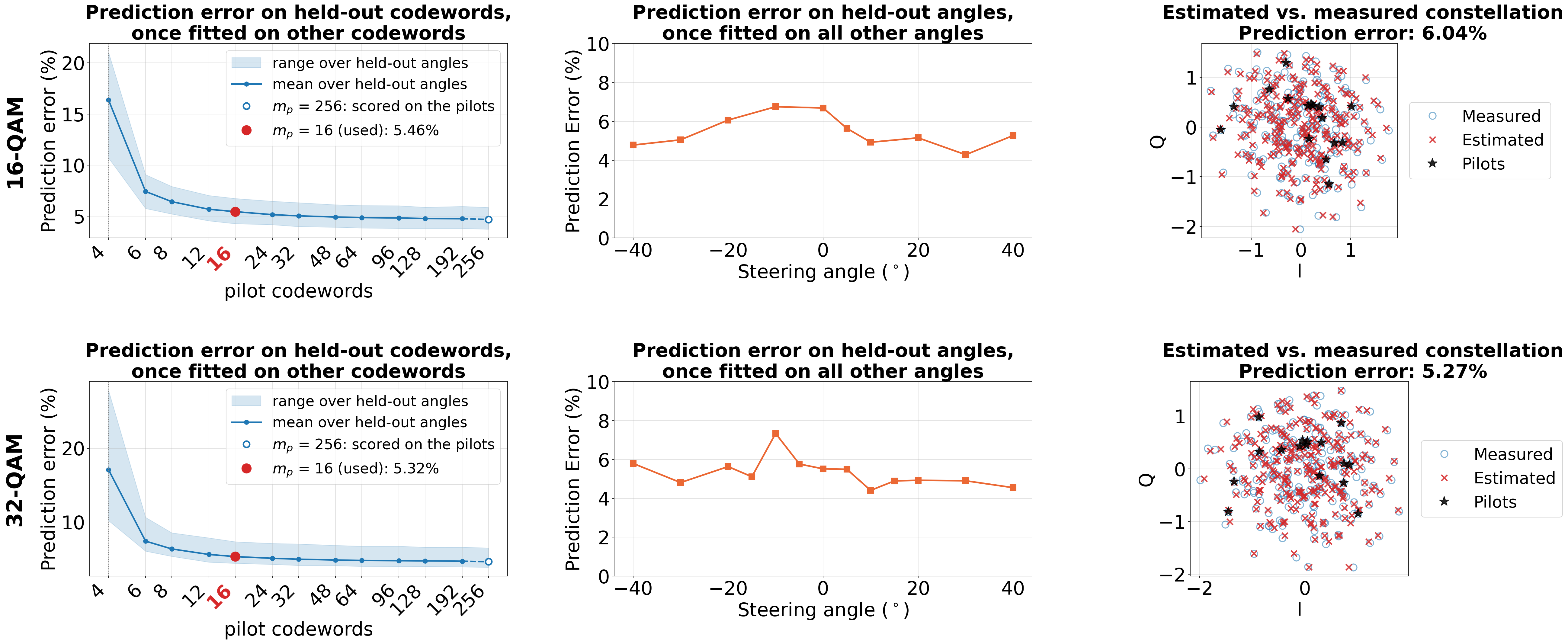}
    \caption{Calibration results for 16- (top) and 32-QAM (bottom) on the 4-channel array.
    From left to right:
    (i) Held-out prediction error versus the number of pilot codewords $m_p$. The line and band are the mean and range over the tested angles. 
    (ii) Prediction error at each held-out steering angle with 16-pilots.
    (iii) Estimated and measured constellations at a held-out angle, $0^\circ$, with 16 pilots (stars) are used to estimate the remaining 240 codewords, and the title gives the prediction error over these 240. Circles mark the measured and crosses the estimated codewords.}
    \label{fig:calibration}
\end{figure*}

\begin{figure}
    \centering
    \includegraphics[width=\linewidth]{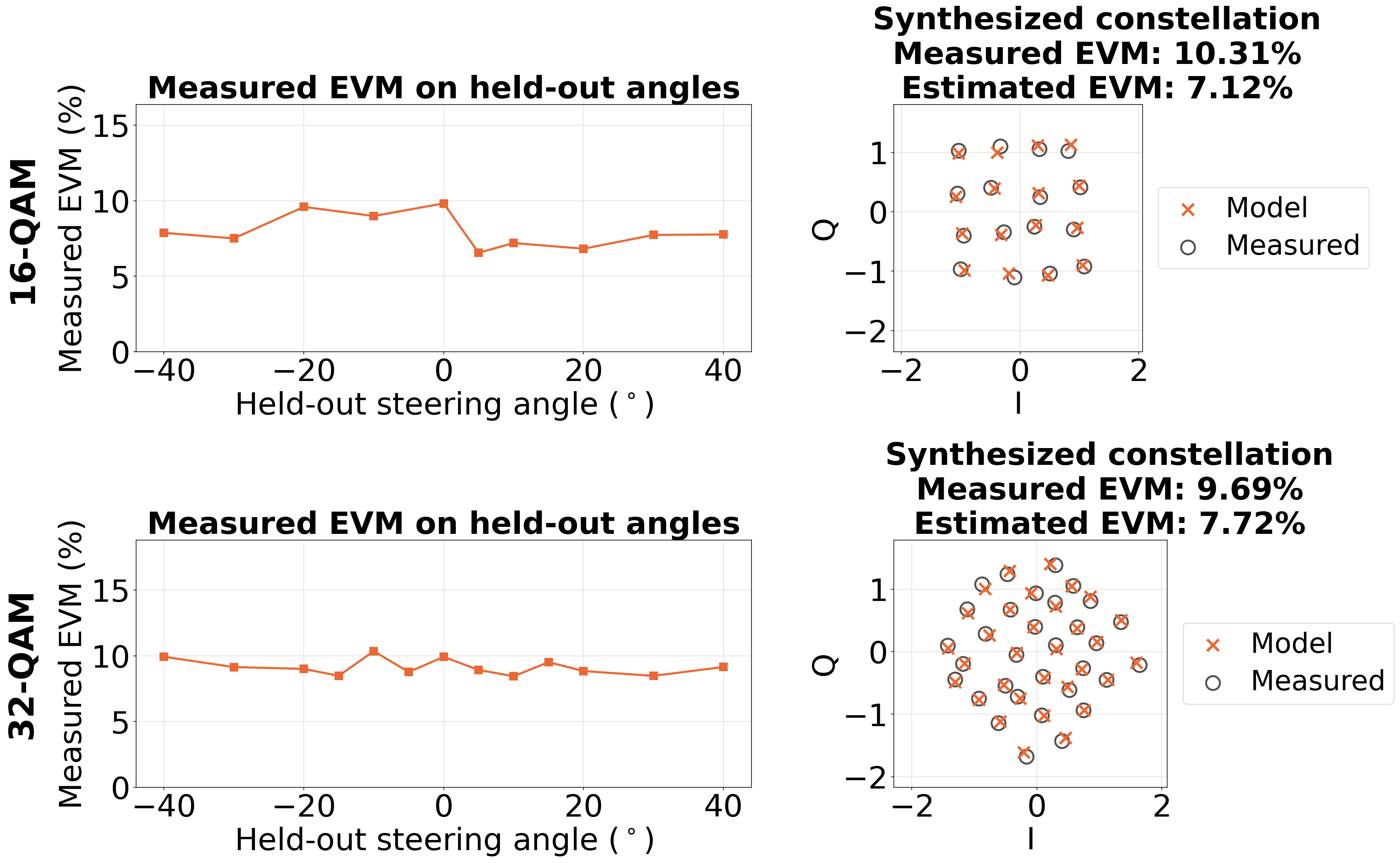}
    \caption{Synthesis results for 16- (top) and 32-QAM (bottom) on the 4-channel array.
    Left: EVM at each held-out steering angle with 16-pilots.
    Right: Measured constellations (circles) and model predictions (crosses) at a held-out angle, $0^\circ$, with 16 pilots.}
    \label{fig:calibration2}
\end{figure}

\subsubsection{Prediction}

Once the shared phase-state gains have been learned at other angles, 16 pilot codewords, i.e., 6.25\% of the codewords, predict the remaining 240 responses at a held-out angle. 
The small pilot budget follows from the model structure. With $\boldsymbol{\Gamma}$ fixed, the only unknowns at a new angle are the $N=4$ complex entries of $\mathbf{a}_e$, determined by $m_p$-pilots.
As shown in Fig.~\ref{fig:calibration}, even when the same model is fitted to all 256 codewords at every angle, errors of 4.62\% (16-QAM) and 4.56\% (32-QAM) remain, and increasing the pilot count to 64 lowers the held-out error only to 4.88\% (16-QAM) and 4.79\% (32-QAM). Therefore, the knee point of $m_p=16$ is chosen.

Table~\ref{tab:model_ablation} reports the error between predicted and measured responses. It shows that the dominant error is from $\mathbf{a}_e$. Fitting $\mathbf{a}_e$ alone lowers the held-out error from 71.29\% (16-QAM) and 66.81\% (32-QAM) to 24.13\% (16-QAM) and 24.67\% (32-QAM), while fitting $\boldsymbol{\Gamma}$ alone raises it to 84.23\% (16-QAM) and 69.90\% (32-QAM). 
That is because $\boldsymbol{\Gamma}$ is shared across angles, while $\mathbf{a}_e$ holds channel-dependent errors of the in-sample angles that do not fully transfer to held-out angles. 
Compared to the ideal case, fitting $\boldsymbol{\Gamma}$ and $\mathbf{a}_e$ provides a reduction of $71.29\% \rightarrow 5.46\%: 13.1\times$ (16-QAM) and $66.81\% \rightarrow 5.32\%: 12.6\times$ (32-QAM). 

\subsubsection{Synthesis}

Calibration carries over into synthesis, as shown in Fig.~\ref{fig:calibration2}. 
With $\boldsymbol{\Gamma}$ learned at the other angles and $\mathbf{a}_e$ from 16 pilots at the held-out angle, the synthesized codebooks have a measured EVM of 7.98\% (16-QAM) and 9.15\% (32-QAM) on average, within 1.7 percentage points of the in-sample reference (6.33\% and 7.68\%).
With the ideal model, the measured EVM is 49.50\% (16-QAM) and 52.56\% (32-QAM) on average, so both $\boldsymbol{\Gamma}$ and $\mathbf{a}_e$ remain necessary for synthesis, as they are for prediction.

\begin{table}[t]
    \centering
    \caption{Held-out prediction error and synthesis EVM on the 4-channel array, averaged over all held-out angles.}
    \label{tab:model_ablation}
    \setlength{\tabcolsep}{4pt}
    \begin{tabular}{@{}l l c c c c@{}}
        \toprule
        & & \multicolumn{2}{c}{Prediction} & \multicolumn{2}{c}{Synthesis} \\
        \cmidrule(lr){3-4}\cmidrule(l){5-6}
        $\boldsymbol{\gamma}$ & $\mathbf{a}_e$ & 16-QAM & 32-QAM & 16-QAM & 32-QAM \\
        \midrule
        Ideal & Ideal & 71.29\% & 66.81\% & 49.50\% & 52.56\% \\
        Fitted & Ideal & 84.23\% & 69.90\% & 77.33\% & 66.20\% \\
        Ideal & Fitted & 24.13\% & 24.67\% & 20.02\% & 21.87\% \\
        \textbf{Fitted} & \textbf{Fitted} & \textbf{5.46\%} & \textbf{5.32\%} & \textbf{7.98\%} & \textbf{9.15\%} \\
        \midrule
        \multicolumn{2}{@{}l}{In-sample} & 4.62\% & 4.56\% & 6.33\% & 7.68\% \\
        \bottomrule
    \end{tabular}
\end{table}

\subsubsection{Drift-awareness}
\label{subsec:hwdrift}

As shown in Table~\ref{tab:hwdrift}, drift-aware synthesis reduces EVM\textsubscript{drift} by $29.44\%\rightarrow 25.06\%: 1.17\times$ (16-QAM) and $29.46\%\rightarrow24.86\%: 1.18\times$ (32-QAM).
EVM at $\theta_s$ increases from 7.86\% to 12.28\% (16-QAM) and 9.12\% to 11.52\% (32-QAM), while EVM at $\theta_{s\pm1}$ decreases from 48.76\% to 38.16\% (16-QAM) and 49.05\% to 39.46\% (32-QAM).

\begin{table}[t]
    \centering
    \caption{Measured drift-aware synthesis on the 4-channel array, averaged over the held-out angles with two equally spaced neighbors.}
    \label{tab:hwdrift}
    \footnotesize
    \begin{tabular}{@{}llcccc@{}}
        \toprule
        Dataset & Design & EVM\textsubscript{drift} & EVM($\theta_s$) & EVM($\theta_{s\pm1}$) \\
        \midrule
        16-QAM & Nominal & 29.44\% & 7.86\% & 48.76\% \\
               & Drift-aware & 25.06\% & 12.28\% & 38.16\% \\ \midrule
        32-QAM & Nominal & 29.46\% & 9.12\% & 49.05\% \\
               & Drift-aware & 24.86\% & 11.52\% & 39.46\% \\
        \bottomrule
    \end{tabular}
\end{table}

\section{Conclusion}
\label{sec:conclusion}

This work formulates ODP modulation as joint spatial and temporal codeword synthesis with channel-dependent amplitude and phase responses and angular sensitivity.
A second-order Taylor approximation of the steering response under Gaussian drift yields a quadratic term of rank at most three. In ideal-hardware simulation, designing with it reduces EVM\textsubscript{drift} by up to 1.54$\times$.

The solver comparison motivates combining greedy coordinate refinement with a randomized start. At 512 channels, $T=1$, and no drift, sidelobe suppression rises from 7.85 to 12.34~dB while EVM remains 4.51\%. This method runs 13--16$\times$ faster than simulated annealing with cached response sums, with sidelobe suppression within 0.7~dB. For four-channel arrays, tiling remains preferable, as it matches exhaustive search.

On the 4-channel 22-nm FD-SOI prototype, phase-state gains learned at other angles transfer to a new angle, where 16 pilots fit only its channel response. Predictions of the remaining 240 responses have mean errors of 5.46\% and 5.32\% for the 16- and 32-QAM datasets, 13.1$\times$ and 12.6$\times$ lower than the ideal-response model.
Codebooks selected from these predictions reach measured EVMs within 1.7 percentage points of selecting from all 256 measured responses.
On the prototype, drift-aware design lowers the measured EVM under drift by up to 1.18$\times$.

\section*{Acknowledgment}

The authors used Claude solely to improve the readability and language of the manuscript. All generated text was carefully reviewed and edited by the authors, who take full responsibility for the content and accuracy of this work.

\bibliographystyle{IEEEtran}
\bibliography{refs}

\end{document}